%% file: main.tex
\documentclass{article}
\usepackage[a4paper, total={6in, 8in}]{geometry}
\usepackage[table]{xcolor}
\usepackage{colortbl}
\usepackage{booktabs}
\usepackage{graphicx}
\usepackage[export]{adjustbox}
\usepackage{csquotes}
\usepackage{multirow, multicol}
\usepackage{tabularx}
\usepackage{pifont}
\usepackage{makecell}
\usepackage{array}
\usepackage{lscape}
\usepackage{hyperref}
\usepackage{xurl}
\usepackage{ifthen}
\usepackage{subfigure}
\usepackage[multiple]{footmisc}
\usepackage{url}
\usepackage{enumitem}
\usepackage{amsmath}
\usepackage{cleveref}
\setlist[itemize]{align=parleft,left=0pt..1em}

\crefformat{section}{\S#2#1#3} %
\crefformat{subsection}{\S#2#1#3}
\crefformat{subsubsection}{\S#2#1#3}

\newcommand{\vcutS}{\vspace*{0cm}}
\newcommand{\vcutM}{\vspace*{0cm}}
\newcommand{\vcutL}{\vspace*{0cm}}
\newcommand{\vcutXL}{\vspace*{05cm}}

\newcounter{mainfindingid}
\newcommand{\mainfinding}[2]{\refstepcounter{mainfindingid}\label{#1}\item[
	\ifthenelse{\value{mainfindingid}=1}{Observation-}{O-}\arabic{mainfindingid}:
	] #2}
\newcommand{\refmainfinding}[1]{\textbf{O-\ref{#1}}}

\newcommand{\customlabel}[1]{\label{#1}}

\newcounter{resquestion}
\newcommand{\resq}[1]{\refstepcounter{resquestion}\item[RQ\theresquestion:\customlabel{resq:#1}]}
\newcommand{\resqref}[1]{RQ\ref{resq:#1}} 

\newcommand{\addition}[1]{{\color{black} #1}}
\newcommand{\modification}[1]{{\color{black} #1}}

\author{Laurens Versluis\footnote{Coresponding author: \href{mailto:l.f.d.versluis@vu.nl}{l.f.d.versluis@vu.nl}}\\
Vrije Universiteit Amsterdam\\
The Netherlands
\and
Mehmet Cetin\\
Vrije Universiteit Amsterdam\\
The Netherlands
\and
Caspar Greeven\\
Surf\\
The Netherlands
\and
Kristian Laursen\\
Vrije Universiteit Amsterdam, Surf\\
The Netherlands
\and
Damian Podareanu\\
Surf\\
The Netherlands
\and
Valeriu Codreanu\\
Surf\\
The Netherlands
\and
Alexandru Uta\\
Leiden University\\
The Netherlands
\and
Alexandru Iosup\\
Vrije Universiteit Amsterdam\\
The Netherlands
}

\title{A Holistic Analysis of Datacenter Operations:\\ Resource Usage, Energy, and Workload Characterization\\ Extended Technical Report}

\date{}

\input{setup}

\begin{document}

\maketitle

\begin{abstract}
	\input{00_abstract}

\end{abstract}

\input{01_introduction}

\input{02_correlation}

\input{03_method}

\input{08_threats_to_validity}

\input{04_metric_characterization}

\input{05_covid}

\input{06_workload_characterization}

\input{07_implications}
\input{09_related_work}

\input{10_conclusion}

\bibliographystyle{plain}
\bibliography{surfing_beautified.bib}

\clearpage

\end{document}

%% file: setup.tex
\usepackage{enumitem}
\usepackage{xspace}
\newcommand{\filler}[1]{\relax}
\definecolor{OwnAzure}{HTML}{336699}
\definecolor{OwnCerulean}{HTML}{CAE2FE}
\definecolor{OwnOliveGreen}{HTML}{556B2F}
\definecolor{KamPurple}{HTML}{907C97}
\usepackage[colorinlistoftodos,prependcaption,textsize=tiny]{todonotes}%
\usepackage{xargs}                      %

\newcommandx{\todolarge}[2][1=]{\todo[inline,size=\large,linecolor=OwnAzure,backgroundcolor=OwnCerulean,bordercolor=OwnAzure,#1]{#2}}

\newcommandx{\todoai}[2][1=]{\todo[inline,linecolor=OwnAzure,backgroundcolor=OwnCerulean,bordercolor=OwnAzure,#1]{#2}}

\newcommandx{\addref}[2][1=]
{\todo[inline,linecolor=blue,backgroundcolor=blue!50,bordercolor=blue,#1]{Add reference. #2}}
\newcommandx{\unsure}[2][1=]{\todo[inline, linecolor=red,backgroundcolor=red!25,bordercolor=red,#1]{#2}}
\newcommandx{\change}[2][1=]{\todo[inline, linecolor=blue,backgroundcolor=blue!25,bordercolor=blue,#1]{#2}}
\newcommandx{\info}[2][1=]{\todo[linecolor=OwnOliveGreen,backgroundcolor=OwnOliveGreen!25,bordercolor=OwnOliveGreen,#1]{#2}}
\newcommandx{\improvement}[2][1=]{\todo[linecolor=Plum,backgroundcolor=Plum!25,bordercolor=Plum,#1]{#2}}
\newcommandx{\thiswillnotshow}[2][1=]{\todo[disable,#1]{#2}}

\definecolor{darkred}{rgb}{0.5,0,0}
\definecolor{darkgreen}{rgb}{0,0.5,0}
\definecolor{darkblue}{rgb}{0,0,0.5}

\graphicspath{{figures/}}

\usepackage{glossaries}

\newacronym{is}{IS}{interest set}

%% file: 00_abstract.tex
Improving datacenter operations is vital for the digital society. 
We posit that doing so requires our community to shift, 
from operational aspects taken in isolation
to holistic analysis of datacenter resources, energy, and workloads.
In turn, this shift will require new analysis methods, and open-access, FAIR datasets with fine temporal and spatial granularity.
We leverage in this work one of the (rare) public datasets providing fine-grained information on datacenter operations. %
Using it, we show strong evidence that fine-grained information reveals new operational aspects.
We then propose a method for holistic analysis of datacenter operations, providing statistical characterization of node, energy, and workload aspects.
We demonstrate the benefits of our holistic analysis method by applying it to the operations of a datacenter infrastructure with over 300 nodes. 
Our analysis reveals both generic and ML-specific aspects, and further details how the operational behavior of the datacenter changed during the 2020 COVID-19 pandemic.
We make over 30 main observations, providing holistic insight into the long-term operation of a large-scale, public scientific infrastructure.
We suggest such observations can help immediately with performance engineering tasks such as predicting future datacenter load, and also long-term with the design %
of datacenter infrastructure.

%% file: 01_introduction.tex
\section{Introduction}
\label{surfing:sct:introduction}

\begin{figure}[t] %
	\centering
	\includegraphics[max width=\linewidth]{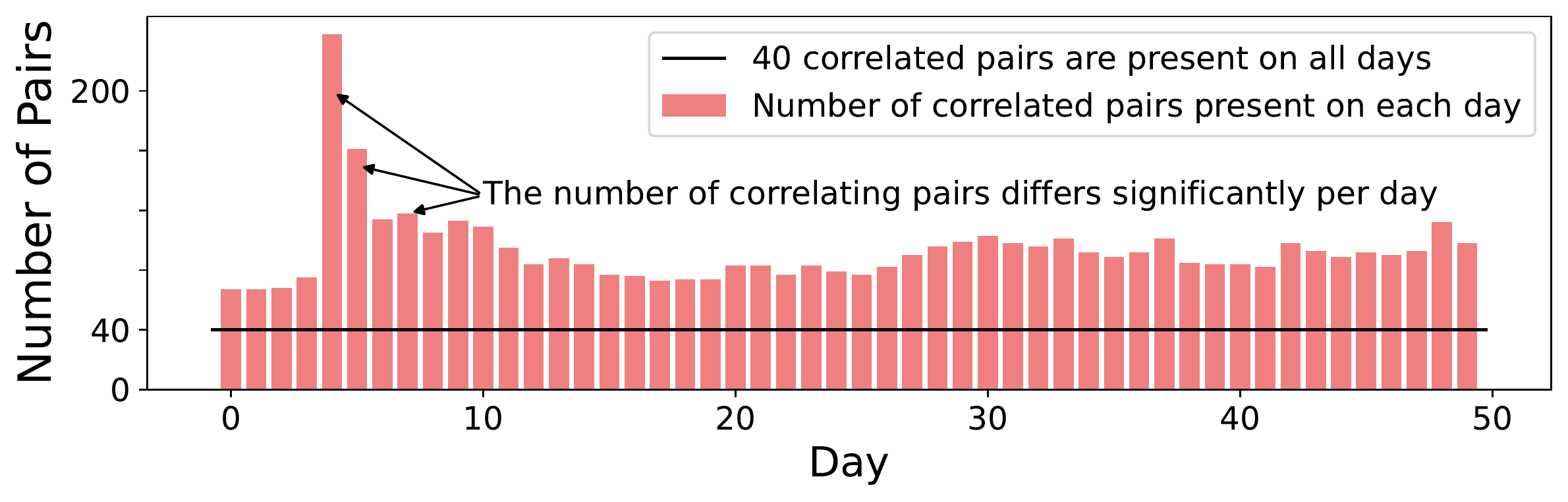}
	\vcutL
	\vcutM
	\caption{Number of correlated metric pairs with a Spearman coefficient $\geq 0.9$ across 50 days. The black line depicts the number of pairs that are consistently present on all days. Because only 40 pairs are consistently correlated, we need to consider the other hundreds of metrics, taken individually. %
	}
	\label{surfing:fig:correlation-metrics-50-days-0.9}
	\vcutL
\end{figure}

Datacenters have \modification{become the main} computing infrastructure for the digital society~\cite{DutchDataCenters20}. 
\modification{Because datacenters are supporting increasingly more users and more sophisticated demands}, their workloads are \modification{changing rapidly.} 
Although our community has much data and knowledge about HPC and supercomputing workloads~\cite{DBLP:journals/jpdc/FeitelsonTK14,DBLP:conf/usenix/AmvrosiadisPGGB18}, 
we have relatively much less information on emerging workloads such as machine-learning, %
which seem to differ significantly from \modification{past workloads}~\cite{mt-ml1,mt-ml3}. %
We posit in this work that, to design the efficient datacenters of tomorrow, we 
need
comprehensive yet low-level machine metrics from datacenters. 
Such metrics could be key to optimization~\cite{lockwood2018year}, performance analysis~\cite{maricq2018taming,uta2020big}, and uncovering important phenomena~\cite{gunawi2018fail}.
Yet, 
\modification{comprehensive datasets of low-level datacenter metrics are rare}~\cite{verma2015large,tirmazi2020borg,DBLP:conf/sosp/CortezBMRFB17,DBLP:conf/usenix/ShahradFGCBCLTR20}. 
\addition{Commercial providers are reluctant to publish such datasets, for reasons that include the need for commercial secrecy, adherence to privacy legislation, and lack of strong incentives to compensate for the additional effort.}
\modification{Often, published datasets are} collected over short periods of time, with coarse time-granularity, not including low-level machine metrics. 
\modification{Some datasets have hardware specifications and rack topologies omitted and values obfuscated through normalization or other processes; only}
coarse, narrowly focused analysis can result from them. 
In contrast, \addition{in this work} we propose a method for holistic analysis of datacenter operations, and apply it to the only \addition{long-term, fine-grained}, open-access dataset~\cite{laursen_olason_kristian_valur_2020_3878143,DBLP:journals/usenix-login/UtaLIMPC20} \addition{that is currently available in the community.} \modification{This dataset contains} long-term and fine-grained operational \modification{server} metrics gathered from a scientific computing infrastructure over a period of nearly 8 months \modification{at 15-second intervals}. %
We show that such data \modification{are} key in understanding datacenter behavior and encourage all datacenter providers to release such data and join our effort.

We focus on addressing three key challenges related to holistic understanding of datacenter operations.
First, the \textit{lack of work on diverse operational metrics.}
For decades, the community has successfully been optimizing computer systems only for the metrics we measured---e.g., throughput~\cite{thain2005distributed}, job completion time~\cite{verma2012two}, latency~\cite{dean2013tail}, fairness~\cite{ghodsi2011dominant}---and biased toward the workloads and behavior that have been open-sourced~\cite{DBLP:journals/jpdc/FeitelsonTK14,DBLP:conf/usenix/AmvrosiadisPGGB18,verma2015large,tirmazi2020borg,DBLP:conf/sosp/CortezBMRFB17,DBLP:conf/usenix/ShahradFGCBCLTR20,DBLP:journals/tpds/VersluisMTHPDI20}. In Figure~\ref{surfing:fig:correlation-metrics-50-days-0.9}, we show evidence motivating the need to capture diverse sets of datacenter metrics. \modification{Using the uniquely fine-grained open-source dataset, we perform an all-to-all correlation analysis on 300+ low-level metrics.
To get a reasonable idea for the number of correlating pairs per day, we investigate 50 separate days.
This number is sufficient to highlight that the number of correlations varies greatly, likely depending on the daily workload of the datacenter.
This suggests that capturing only a few metrics is insufficient to get a comprehensive view on the datacenter operation, as most metrics cannot be reliably derived from another.} %
Instead, to capture possibly vital information, we should aim to include 
as much data as possible, from hardware sensors, to operating systems, and further to the application level. 
\addition{(In Section~\ref{surfing:sec:implications:overhead}, we assess the additional storage costs for fine-grained data sampling at 15-second intervals and, together with the datacenter operators that published the open-access dataset, we interpret the results as indicative that these costs are worthwhile for the additional insight they enable.)}

\addition{We identify as the second main challenge}
the \textit{lack of holistic analysis methods}, able to \addition{combine and then} work on diverse operational metrics \modification{such as workload and machine metrics.}
Previous research already points out that large bodies of modern research might be biased toward %
the
available datasets~\cite{DBLP:journals/cacm/BouwersVD12,DBLP:conf/usenix/AmvrosiadisPGGB18}, and that effort to measure ``one level deeper'' is still missing~\cite{DBLP:journals/cacm/Ousterhout18}.
Next to operational bias, this also results in understudying other metrics and limits our ability to fully understand large-scale computer systems.
For example, only since the 2000s and more intensively only after the mid-2010s, has energy consumption become a focus point~\cite{dayarathna2015data,goiri2011greenslot}.
In pioneering work in operational data analytics in the late-2010s, Bourassa et al.~\cite{DBLP:conf/icppw/BourassaJBCJVS19} propose to conduct extensive data collection and feed the results back into running datacenters for improving operations.
Pioneering software \modification{infrastructures} such as GUIDE~\cite{DBLP:conf/sc/VazhkudaiMTZWOG17} and DCDB Wintermute~\cite{DBLP:conf/hpdc/NettiMGOTO020} take first steps in this direction. 
However, much more research is needed to understand the kinds of data and analysis feasible (and necessary) in this field.
Similarly, many studies and available datasets focus only on computational aspects, e.g., \cite{DBLP:conf/usenix/AmvrosiadisPGGB18, DBLP:journals/tpds/VersluisMTHPDI20, DBLP:conf/eScience/SilvaCJVD14, DBLP:journals/jpdc/FeitelsonTK14},
but 
details on the operation of machine-learning workloads on infrastructure equipped with GPUs (and, further, TPUs, FPGAs, and ASICs) are still scarce.

\addition{As the third main challenge, we consider}
the relative \textit{lack of relevant, fine-grained, and public datasets}.
In practice, collecting holistic data has been feasible at the scale of datacenters for nearly a decade, with distributed monitoring~\cite{DBLP:journals/cacm/LegrandVCGBC09}, tracing~\cite{2010-dapper}, and profiling~\cite{2016-osdi-profiling} tools already being used in large-scale datacenters. 
Unfortunately, such data rarely leaves the premises of the datacenter operator. 
From the relatively few traces that are shared publicly, many are focused on important but specific kinds of workloads, such as tightly-coupled parallel jobs~\cite{DBLP:journals/jpdc/FeitelsonTK14}, bags of tasks~\cite{DBLP:journals/fgcs/IosupLJADWE08}, and workflows~\cite{DBLP:journals/tpds/VersluisMTHPDI20}. %
Other datasets only include a limited subset of metrics such as power consumption~\cite{DBLP:conf/ipps/PatelWEHZT20}, or high-level job information~\cite{DBLP:conf/sc/PatelLKRAT20}.
Only a handful of datasets include low-level server metrics, such as the Microsoft Azure serverless traces~\cite{DBLP:conf/usenix/ShahradFGCBCLTR20} or the Solvinity business-critical traces~\cite{DBLP:conf/ccgrid/ShenBI15}.
Recently, in 2020, the largest public infrastructure for scientific computing in the Netherlands has released as
\addition{Findable, Accessible, Interoperable, Reusable~(FAIR)}~\cite{data:FAIR16}, open-access data a long-term, fine-grained dataset about their operations~\cite{laursen_olason_kristian_valur_2020_3878143,DBLP:journals/usenix-login/UtaLIMPC20}. %
In this work, \textit{we take on the challenge of conducting the first holistic analysis of the datacenter operations captured by this dataset}.

Addressing these challenges, we advocate for 
a holistic view of datacenter operations, with a four-fold contribution:
\begin{enumerate}[leftmargin=*]
    \item We analyze whether diverse operational metrics are actually needed~(Section~\ref{surfing:sct:correlations}). 
    We conduct a pair-wise correlation study across hundreds of server metrics, and analyze whether correlations are enough to capture datacenter behavior. Our results show strong evidence about the need for a more diverse set of metrics, to capture existing operational aspects.
    
    \item Motivated by the need for diverse operational metrics, we propose a holistic method for the analysis of datacenter operations~(Section~\ref{surfing:sec:method}).
    Our method considers information about machine usage, energy consumption, and incoming workload, and provides comprehensive statistical results.
    
    \item We show the benefits of our method in understanding the long-term operations of a large public provider of scientific infrastructure~(Sections~\ref{surfing:sec:machines}--\ref{surfing:sec:workload}).
    We provide the first holistic insights into a large-scale, fine-grained and, most importantly, public     dataset~\cite{laursen_olason_kristian_valur_2020_3878143,DBLP:journals/usenix-login/UtaLIMPC20}---over 60\,billion data-points collected over 7.5 months at the high frequency of 15\,seconds. %
    Unique features %
    include a comparison of generic and machine-learning workloads and nodes, per-node analysis of power consumption and temperature, and glimpses at the COVID-19 period.
    
    \item We explore ways to leverage holistic-analysis results to improve datacenter operation~(Section~\ref{surfing:sec:implications}).
    We first investigate short-term use, %
    quantifying how higher-frequency information leads to better %
    online load prediction. 
    We propose actionable insights, assessing overheads of collecting more data and metric correlations.
    We also exemplify long-term use for design and tuning. %
    
\end{enumerate}

%% file: 02_correlation.tex
\section{Are Just a Few Metrics Enough?}
\label{surfing:sct:correlations}
\label{surfing:sec:correlations}

\begin{figure*}[t]
    \centering
    \includegraphics[max width=\textwidth]{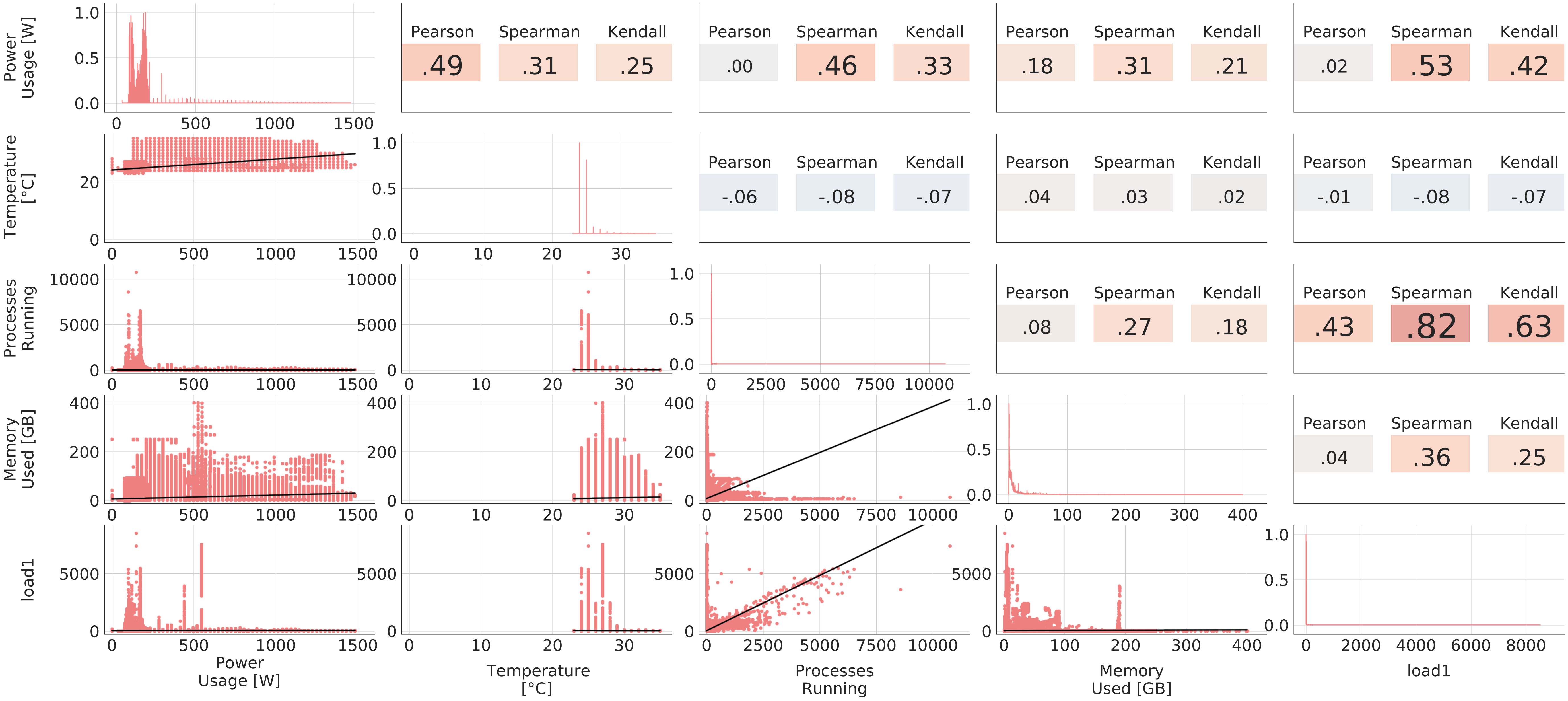}
    \vspace{-5cm}
    \vcutXL
    \caption{The correlation matrix of node metrics. See text for an explanation of the sub-plots.} %
    \label{surfing:fig:correlation-matrix-node-metrics}
\end{figure*}

In this section we show that more metrics are needed when analyzing datacenter behavior, and thus also that more metrics should be recorded and shared. We reach this conclusion by analyzing correlations for a rich set of over 300 low-level metrics collected from a scientific and engineering datacenter~(details in Section~\ref{surfing:sec:method:input}). 
Although the dataset includes low-level metrics collected by servers, OS, and applications, we focus in this section on metrics mostly as context-agnostic information, that is, without a structure or ontology that attaches them to specific datacenter components or processes. This allows us to understand whether more metrics can provide new information. 
(We address metrics in their context in our analysis method, in the next section.)

\textbf{Method overview:}
Correlations can lead to improvements in system monitoring and find interesting relationships for, e.g., predictions~\cite{DBLP:journals/concurrency/XiaoYER16}.
In particular, we are interested if all metrics in the dataset we consider 
are necessary or can be obtained from others through correlation, and if these correlations are persistent or workload-dependent.
First, we compute all valid correlation-pairs during a day and inspect if pairs which are considered \enquote{very strong} by literature are persistent, as these pairs are the most likely candidates to reduce the size of the dataset through derivation and likely most robust. Second, we analyze visually the distribution and correlations of several commonly used high-level node metrics.

\textbf{Conclusion:}
A small set of metrics cannot capture the information provided by diverse metrics. 
We urge datacenter practitioners to \textit{collect as much fine-grained data as possible for enabling valuable analyses}, and 
to \textit{open-source such data for the benefit of all}.

\subsection{More Metrics Needed}
\label{surfing:sec:correlations:moremetricsneeded}

To observe if metrics are workload dependent, we compute the Pearson, Spearman, and Kendall correlations for all metric pairs, for 50 different days.
This results in over 14,000 valid correlation pairs per day\footnote{For some metrics, no valid data is available or all values are the same.}.
Next, we compute per day the number of pairs with \enquote{very strong} correlation, i.e., with Spearman coefficient $\geq$0.9~\cite{schober2018correlation}.
To verify that all coefficients are significant, we verify the probability of an uncorrelated system producing a correlation as extreme as in our dataset is negligible, i.e., all p-values of the pairs depicted in the figure are equal to 0\footnote{\url{docs.scipy.org/doc/scipy/reference/generated/scipy.stats.spearmanr.html}}.

Figure~\ref{surfing:fig:correlation-metrics-50-days-0.9} depicts the number of correlated metric-pairs, per day. 
We observe the number of pairs fluctuates significantly, with only 40 pairs present on all days.
This indicates that correlations, even very strong ones, change daily---because workloads are the most variable aspect of a datacenter, we conjecture correlations are workload-dependent. This suggests metric information should be collected across many metrics, and over long periods of time.
Second, this shows, combined with observations from
Section~\ref{surfing:sec:correlations:notenough}, 
that we cannot (significantly) reduce the amount of metrics, as many of these metrics cannot be reliably derived nor predicted from others.

\subsection{Correlations Are Not Enough}
\label{surfing:ssct:node-metric-correlation}
\label{surfing:sec:correlations:notenough}

To understand what correlations can reveal about the correlated metric-pairs, we depict in Figure~\ref{surfing:fig:correlation-matrix-node-metrics} a correlation matrix of power usage, ambient temperature, number of processes running, amount of memory used, and UNIX load1. 
For all these metrics, the input dataset has valid data, so we are able to accurately compute all correlations.

The correlation matrix in Figure~\ref{surfing:fig:correlation-matrix-node-metrics} includes: 
(i) distribution plots on the diagonal,
(ii) pair-wise scatter plots and linear regressions in its sub-diagonal elements; 
(iii) pair-wise Pearson, Spearman, and Kendall correlations between metrics, in mirrored elements of (i).

Based on (i), we observe most metrics have a long tail.
We also observe that the majority of values for temperature and somewhat for memory used is confined to a constrained area.
From (ii), we observe that most combinations of metrics do not seem to have a linear relationship.
Four pairs of metrics seem to have either some or a strong linear correlation.
If we look at the Pearson, Spearman, and Kendall correlations corresponding to figures in (ii), we observe some additional insights going by the suggestion of Schober et al. to use these correlations as \enquote{a measure of the strength of the relationship}~\cite{schober2018correlation}. (All p-values are $<10^{-13}$, so the results are meaningful.)

load1 and power usage seem to correlate somewhat going by the Spearman correlation which does not show in (ii), and second, the reverse shows for memory usage and number of running processes where relation in the regplot does not appear in any of the correlations.
Even though one might assume a (seemingly) linear regression would also show in a ranking correlation such as Spearman, this does not always hold.

The temperature seems to correlate somewhat with the power usage of a node.
This makes sense as initially as power usage increases, the temperature goes up due to heat dissipation.
Naturally, the temperature eventually starts to stabilize and even goes down as the components within the system start to get cooled by the cooling system.
The second pair that shows some relationship is memory used and number of processes running.
The scatterplot shows a linear regression curve, yet as the data is not normal distributed, the Pearson correlation is close to 0. 
The Spearman correlation shows a moderate relationship between these two metrics. 
As can be observed from the regression curve, the amount of memory used does go up with the number of processes running, yet many outliers exist.
The scatter plot shows that sometimes the amount of memory used in the system reaches the maximum when the number of processes running observed is low.
We also observe that there are many measurements where the number of processes running is high, yet the total memory used by the node remains low, indicating that whenever many processes are running, they are lightweight in terms of memory usage.
The third pair that shows a strong correlation is load1 and the number of processes running, with a Spearman coefficient of 0.82 and Kendall coefficient of 0.63.
The load1 metric roughly depicts the average number of active processes in the last minute, which naturally should correlate well with the number of processes running.
As the regression curve almost has a slope of 1, this indicates that bursts are infrequent; as the load1 is an average over the past minute and processes running is real-time, bursts may be dampened if they are very short running processes.
We observe that many measurements are just below this curve, i.e., the number of processes running is higher than load1. 
This indicates that some processes are not awaiting resources and are e.g., suspended.
We also observe some measurements where the load is higher than the number of processes running, this indicates a possible burst of short running processes that causes the average of load1 to spike, yet do not reflect in the current number of processes running.
The other pairs of metrics do not seem to correlate, indicating that these combinations can't be used to predict a counterpart.
This highlights the complexity of these systems and the difficulty in understanding the parameter space on a system's behavior.

%% file: 03_method.tex
\section{Method for Holistic Analysis}
\label{surfing:sct:method} \label{surfing:sec:method}

We propose in this section a holistic analytical method for datacenter operations. 
In our method, obtaining a holistic view requires combining machine, energy, and workload data; doing so with long-term and fine-grained data enables meaningful findings. 
Our method is data-driven, and thus we address the input data~(Section~\ref{surfing:sec:method:input}) and its cleanup~(Section~\ref{surfing:sec:method:cleanup}).
The highlight of this section is the data analysis, for which we describe the main research questions and how we address them~(Section~\ref{surfing:sec:method:analysis}).
We also cover in this section the practical aspects~(Section~\ref{surfing:sec:method:software}), e.g., the software and our provisions to ensure the reproducibility of this work, and the main limitations we see to our method~(Section~\ref{surfing:sec:limitations}).

\subsection{Input Data} \label{surfing:ssct:cluster-outline} \label{surfing:sec:infrastructure} \label{surfing:sec:method:collection}
\label{surfing:sec:method:input}

Although our method does not depend on specific metrics, we are mindful of the information currently available as public datasets.
We take as model the public dataset with the finest temporal and spatial granularity available---
a dataset open-sourced by %
the largest public provider of scientific infrastructure in the Netherlands. 
The datacenter operators have shared low-level server metrics collected every 15\,seconds for a period of nearly 8\,months~\cite{laursen_olason_kristian_valur_2020_3878143,DBLP:journals/usenix-login/UtaLIMPC20}. %

\textbf{Overall:} Table~\ref{surfing:tbl:generic-outline-trace} summarizes the public dataset: up to 1,26\,million samples per metric per node, and in total 66\,billion individual, high- and low-level metric measurements. The \textit{low-level metrics} include server-level (e.g., power consumption), hardware-sensor (e.g., fan speeds, temperature), and OS-level metrics (e.g., system load).

\textbf{Workload:} The datacenter acts as infrastructure for over 800 users, who have submitted in the period captured by the dataset over 1 million jobs. 
The majority of these jobs originate from the bioinformatics, physics, computer science, chemistry, and machine learning domains. Jobs are exclusive per user; there are no multi-user jobs or workflows at the moment.
SLURM is the cluster manager used to allow users to queue jobs for these different types of nodes.
All jobs are scheduled using FIFO per stakeholder, with fairsharing across stakeholders.
Through the use of queues, the datacenter offers both co-allocation of jobs on the same node and reserving of nodes for exclusive use.
The operator uses \texttt{cgroups} to enforce CPU and memory limits on multi-tenant nodes.

\textbf{Infrastructure:} In total, the datacenter contains 341 nodes spread across 20 racks. 
Racks are either \textit{generic}, including nodes only with CPUs, or for machine learning~(\textit{ML}), including both CPUs and a number of GPUs per node. 
Over 90\% of the workload on the GPU nodes is from the ML domain, a determination based on the libraries used by each job and later checked by the datacenter administrators
Each rack includes up to 32~\textit{generic nodes} or up to 7\,\textit{ML nodes}; the counts depend on GPU types and on power-consumption limitations imposed by the cooling system.

\begin{table}[!t]
	\caption{Generic outline of the machine metric dataset.}
	\label{surfing:tbl:generic-outline-trace}
	\vcutM
	\adjustbox{max width=\linewidth}{
	\begin{tabular}{llr}
		\toprule
		Dataset & Item                                                                             & Value               \\ \midrule
        Public data & Start date                                                                       & 2019-12-29\\
		(see $\S$\ref{surfing:sec:method:input}) & End date                                                                         & 2020-08-07\\
		& Sampling frequency [s] & 15\\
		& \makecell[l]{Max. samples per metric per node} & 1,258,646            \\
		& Number of metrics                                                                & 327                 \\ 
		& Number of measurements &  66,541,895,243 \\
		\midrule
		Clean data & Number of valid racks    & 15 \\
		(see $\S$\ref{surfing:sec:method:cleanup}) & Number of valid nodes & 315 \\
		 & Number of valid measurements &  63,978,689,791 \\
		\bottomrule
	\end{tabular}
}
	\vcutM
\end{table}

\subsection{Data Cleanup}  \label{surfing:sec:method:cleanup}

After inspecting the data, 
we inquired with the dataset creators about (in)valid and missing data, 
and, finally, created cleanup scripts. %
After careful data-cleanup, the dataset we use in this work is unprecedentedly rich, covering the operation of 15\,racks containing 315\,nodes with nearly 64\,billion measurements, spanning over 7\,months.
To clean the dataset in Table~\ref{surfing:tbl:generic-outline-trace}, we focus on:

\textbf{Clean node- and rack-data:} We include only the 315 nodes in 15 racks that are used for computation.
Together, these nodes contain 5,352\,CPU cores, 41.6\,TB of CPU memory, 128\,GPUs, and 1.8\,TB of GPU memory.
Most nodes~(283) only contain CPUs; the others~(32) also have GPUs attached.  

\textbf{Clean job-data:} For the workload, we filter out the jobs based on their start time if they are outside the start and end time range of the dataset.
Additionally, all jobs that are not related to the racks in the machine dataset are filtered out.
These jobs originate from nodes in the 5 racks used as gateways for the public, as debug and testing resources, and as compile farms. %

\textbf{Clean metric-data:} When performing numerical analyses, we removed the NaN values or set them to, e.g., zero when summing.
Overall, the original dataset contains over 66 billion measurements, with close to 2.6\,billion NaN values~(3,85\%).
For some metrics, the dataset contains some gaps where the monitoring system was down; %
for some others, data collection stopped halfway into May 2020.

\textbf{Clean time-series:} %
We filter out all missing measurements (not-a-numbers, NaNs). %
In visual overviews, we mark missing data using special coloring.

\textbf{Clean correlation-data:} When computing correlations between pairs of metrics, we omit pairs where one or both metrics' measurements never change, because such data is unfit for the ranking step required to compute the Spearman and Kendall correlations.

\begin{figure*}[!t]
	\centering
	\includegraphics[max width=\linewidth]{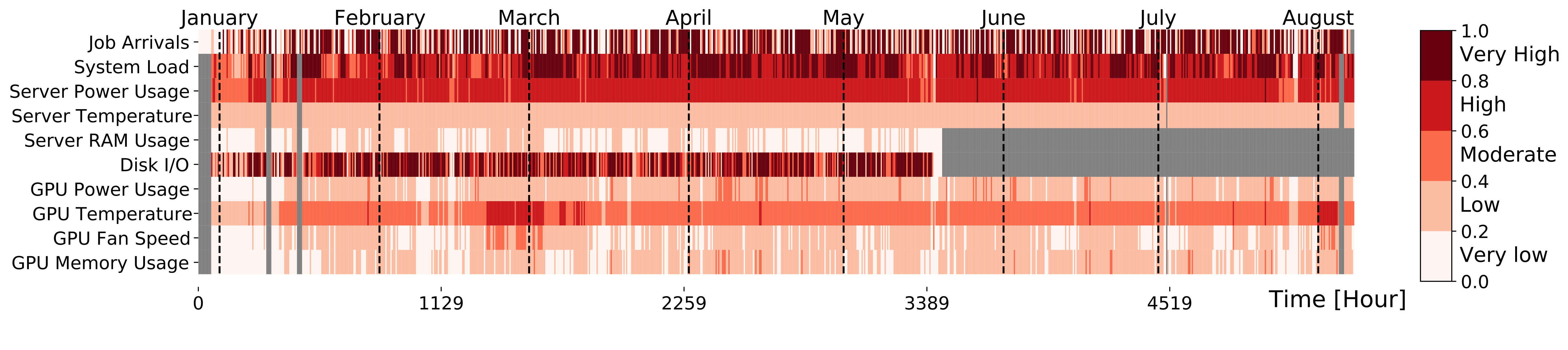}
	\vspace*{-0.25cm}
	\vcutL
	\vcutM
	\caption{Resource usage for various metrics. For this plot, we normalize the metrics and color them accordingly (see text). Vertical dashed lines depict the start of a month. Grey depicts lack of valid data (see \cref{surfing:sec:method:cleanup}).}
	\label{surfing:fig:characterizations:general-resource-usage}
	\vcutM
\end{figure*}

\subsection{Data Analysis}  \label{surfing:sec:method:analysis}

Our method for holistic analysis proposes diverse research questions, %
answered using
fine-grained machine %
and %
workload data.

\modification{\textbf{Machine and workload data:} As main input dataset, we use the clean dataset introduced in Section~\ref{surfing:sec:method:cleanup}. 
For the COVID-19 analysis, we record that the Dutch government declared the start of the (ongoing) pandemic on Feb 27, 2020~\cite{corona}; we thus consider all data before this date to be \enquote{non-covid} data. %
For the workload analysis, %
\modification{the datacenter cannot publish the workload data} due to privacy constraints~(the EU~GDPR law); 
instead, 
we contacted the datacenter operators and worked with them to run the analysis we need on the data. %
}

\addition{\textbf{Method FAIRness~\cite{data:FAIR16}:} 
The scientific community is a powerful advocate for 
FAIR data.
The dataset used in this work is FAIRly stewarded by Zenodo, and comes with a full specification and a data-schema that allow sharing and using the data with low effort~\cite{DBLP:journals/usenix-login/UtaLIMPC20}.} %

\textbf{Novelty of our method:} Previous work~\cite{DBLP:conf/usenix/AmvrosiadisPGGB18,DBLP:conf/sc/VazhkudaiMTZWOG17,DBLP:conf/hpdc/NettiMGOTO020, DBLP:conf/sc/PatelLKRAT20} has performed individual analyses that align and overlap with our holistic analysis. However, the kind of analysis we propose in this work is novel through its all-encompassing scope and detail of the data: we analyze workload (e.g., jobs) data and fine-grained machine data, and show that this is needed to better understand job-machine interaction and to perform predictions. 
We present %
three types of research questions (RQ) addressed by our novel analysis and \textit{mark with a star~($\star$) the RQs which are not answered in any prior work}:

\noindent\textbf{A. Analysis of machine operations} (results in Section~\ref{surfing:sec:machines}): 
To analyze how the datacenter machines behave over a long period of time, 
we use a variety of low-level metrics
as input for answering the following questions:
\begin{description}[leftmargin=0.3cm]

    \resq{resource:usage} \textbf{What is the general resource usage?} 
    We aim to understand the usage of each server: the average system load; RAM, disk I/O, and GPU usage. We further study the average power consumption, the temperature, and the fan speed. %
    
    \resq{resource:memio} \textbf{What is the specific memory and network usage?} 
    The answer should include common ranges and modes in the distribution of memory consumption, etc., per node-measurement;
    linked when possible to known workload.
    
    \resq{resource:power}$\star$ \textbf{What is the power consumption, per node and per rack? What is the rack temperature?} 
    We seek the (instantaneous) power consumption, including common ranges and modes. 
    We want to further understand how the heat dissipates and if the cooling system is overwhelmed.
    
    \resq{resource:diurnal} \textbf{How does the system load vary over time?} 
    We focus here on diurnal and longer-term patterns. 
    (The current dataset does not enable seasonality analysis, but data keeps accumulating.)
    
    \resq{resource:ml}$\star$ \textbf{How do generic and ML nodes and racks differ?}--orthogonal concern, applies to all other machine-related question. 
    
    \resq{wl:covid}$\star$ \textbf{What is the impact of the COVID-19 pandemic?}, especially how operations responded to workload changes.
    
\end{description}

\noindent\textbf{B. Analysis of datacenter workload} (results in Section~\ref{surfing:sec:workload}): 
To understand if the \textit{workload} exhibits similar properties to other traces known in the community, especially traces from scientific and Big Tech clusters, we formulate the following questions:
\begin{description}[leftmargin=0.3cm]

    \resq{wl:jobchars} \textbf{What are the job characteristics?}---job size in CPU-cores, job length, and variability across these features.
    
    \resq{wl:arrival} \textbf{What are the job arrival patterns?} 
    This question focuses on the basic statistics and time-patterns of job submissions. %
    
    \resq{wl:peak} \textbf{What is the peak demand?}---explains the intensity of the peak demand, and contrasts it to normal operation. 
    
    \resq{wl:failure} \textbf{What are the patterns of job-failure?}---fraction of jobs fail to complete and their resource-waste.
    
    \resq{wl:long} \textbf{How do long jobs behave?} 
    We consider this orthogonal concern for each of the other workload-related questions. %
    
\end{description}

\noindent\textbf{C. Generating insights from data} (results in Section~\ref{surfing:sec:implications}): 
\begin{description}[leftmargin=0.3cm]

    \resq{wl:lstm}$\star$ \textbf{How can we leverage fine-grained data?}, focusing on using fine-grained data to perform better predictions.
    
      \resq{wl:storagecomp}$\star$ \textbf{What are the implications of storing fine-grained data?} 
    This question focuses on the feasibility of storage for fine-grained metric data as well as how scalable its analysis is.
    
     \resq{wl:corr}$\star$ \textbf{How do metrics correlate?} 
    This question focuses on insights into low-level metrics correlation and the implication for data collection and analysis.
    
      \resq{wl:implications}$\star$ \textbf{What are the implications of holistic analysis for datacenter operation and design?} 
    This %
    focuses on leveraging fine-grained data to tune and design efficient datacenters. %
    
\end{description}

\subsection{Software and Reproducibility}
\label{surfing:sec:method:software} \label{surfing:sec:method:repro}

To enable reproducibility, we validate and open-source all the software (scripts) used in this work. 
All scripts are checked for correctness by at least two persons.
They load raw data from the dataset available as FAIR~\cite{data:FAIR16}, open-access data at:\\ 
\indent\url{https://doi.org/10.5281/zenodo.4459519}

In our analysis, 
we use Pandas 1.2.0, NumPy 1.19.4, SciPy 1.5.3, and statsmodels 0.12.1.
For some analyses, we use a distributed version of Pandas, Koalas 1.5.0, deployed on a spark cluster running Spark 3.0.0 with Hadoop 2.7.7.
All our analyses and plotting code is available open-source at \url{https://github.com/sara-nl/SURFace}.

%% file: 08_threats_to_validity.tex
\subsection{Known Limitations}
\label{surfing:sct:threats-to-validity}
\label{surfing:sec:limitations}

We discuss here \modification{four} known limitations to our method:

The most important limitation to our method derives from its \textit{holistic nature}, which is also its strength. 
This nature is reflected in the broad analysis of several hundred metrics, which, as we show in the next three sections, %
helps understand how the whole works and gives actionable insights. 
However, datacenters can expose thousands of signals, so even our broad selection imposes a bias. %
Finding a \textit{complete and general}, holistic method of analysis is beyond the scope of this work---a goal which we envision for the entire community, for the next decade, which already includes award-winning work that focuses on selecting meaningful signals~\cite{DBLP:conf/wosp/XiongPZG13} and large-scale data collection~\cite{DBLP:conf/sc/VazhkudaiMTZWOG17,DBLP:conf/ipps/PatelWEHZT20,DBLP:conf/hpdc/NettiMGOTO020}. 
Furthermore, the method proposed here can be contrasted with methods from the other end of the holistic-reductionist spectrum; \textit{compared with focused work} on even one of the questions we address, our method cannot produce the same depth for the same effort. 
Without rehashing the broad and as-of-yet inconclusive debate of the entire scientific world 
about holism vs. reductionism, we draw attention to its current stand-off: both add value and should not be discarded, lest the community that does so fails in producing scientific discoveries, long-term.

A second
limitation
derives from the \textit{statistical methods} used in this work and from the libraries that compute them.
\addition{We use linear regression, because this is the most common form of fitting and thus it is likely to be understood by every member of the community. However, we envision that expert-level models could be developed, e.g., leveraging machine learning or higher order polynomials, giving better accuracy and precision. An example here could be to develop non-linear models where failures and even performance anomalies~\cite{DBLP:journals/csur/IbidunmoyeHE15} are causally linked to signals from many metrics in the system, such as high load, extreme temperature, or unusual~\cite{DBLP:conf/ispdc/GhiasvandC19} and/or fail-slow hardware failures~\cite{gunawi2018fail}.}
As discussed in Section~\ref{surfing:sct:correlations}, %
most metrics are not uniformly distributed, which is required for the Pearson correlation; nonetheless, %
the three correlation coefficients sketch a better picture together. %

Another limitation
is the \textit{vantage point}, in that we look at data from a specific datacenter. This could affect especially the workload-level, where machine learning is emerging. %
However, more datasets as fine-grained as this work analyzes are currently not available publicly---we encourage datacenter operators to help!

\addition{
Last, the dataset we analyze is much more fine-grained than others, but there is still much room for additional data and further analysis of it. 
For example, 
datasets could further include details on 
(i) the \textit{operational policies}, e.g., detailed scheduling queues and thresholds (e.g., in the Parallel Workloads Archive, as defined by the community since the late-1990s~\cite{DBLP:conf/ipps/ChapinCFJLSST99}); 
(ii) the \textit{trade-offs} considered during the capacity planning and datacenter design phases (e.g., of  capability and cost); and 
(iii) the \textit{energy sourcing and flows} (e.g., how the datacenter operations link with the energy markets and renewable energy-generation processes).
}

%% file: 04_metric_characterization.tex
\section{Datacenter Machine Operations}
\label{surfing:sct:machine-metric-characterization}
\label{surfing:sec:machines}

We present in this section a comprehensive characterization of machine operations in datacenters, with the method from Section~\ref{surfing:sec:method:analysis}.

\subsection{General Resource Usage}
\label{surfing:ssct:general-resource-usage}

\begin{description}
	\mainfinding{surfing:mf:queued}{Job arrivals do not consistently overlap with machine metrics, including load, disk I/O. Jobs get queued.}
	\mainfinding{surfing:mf:lisa-high-load}{Average system load is high (44.6\%) or very high (20.2\%).}
	\mainfinding{surfing:mf:lisa-low-ram-usage}{Average RAM usage is low (33.3\%) or very low (66.7\%).}
	\mainfinding{surfing:mf:lisa-lots-of-disk-io}{Average Disk I/O activity is high (1.3\%) or very high (0.8\%).}
	\mainfinding{surfing:mf:lisa-gpu-usage-low}{GPU metrics indicate low (12.5\%-64.4\%) to moderate (1.0\%-80.7\%) average GPU usage.}
\end{description}

To obtain a holistic view of the workload and how resources are being used, we plot the number of jobs arriving and various resource-related metrics in Figure~\ref{surfing:fig:characterizations:general-resource-usage}.
Each slice of a bar in the figure depicts an hour, where the color of the given slice is set to maximum normalized value observed within that hour.
For the arrival of jobs, we count how many jobs arrive per 15 second interval (aligned with the metric samples) and then normalize the data using the 99$^{th}$ percentile and clip the values to 1.
We use the 99$^{th}$ percentile to avoid that a few outliers skew the normalization. %
We then label five intensity classes---very low, low, moderate, high, and very high---, spread equally in the normal range, [0,1].

\textbf{Setup:} To depict how the overall datacenter is utilized, we use UNIX \textit{load1} as system load metric. UNIX load captures the "number of threads that are working or waiting to work"~\cite{BrendanGreggLoad1}. %
The load is an aggregate metric over time, e.g., load1 uses a one minute rolling window.
The load can exceed the number of available server cores indicating the system is likely overloaded. We sum the load1 across all nodes and divide this number by the total available cores within the cluster, clipped to 1 as the values can reach well above 1, since there can be many more threads/processes running or queueing than available cores. 

Further, we show the average \textit{server power usage} normalized to 5,500 Watts which is the maximum the cooling system can handle per rack.
The \textit{server temperature} is normalized to the minimum of the maximum allowed temperatures for the different CPU models, which is the Intel® Xeon® Silver 4110 Processor having a limit of 77 degrees Celsius\footnote{\url{https://ark.intel.com/content/www/us/en/ark/products/123547/intel-xeon-silver-4110-processor-11m-cache-2-10-ghz.html}}.

The \textit{Server RAM usage} shows the utilization of all the RAM in the datacenter. %
To obtain \textit{disk I/O} usage, we sum the bytes read and written \addition{from both local storage and NFS mounts} and divide \addition{this number} by the peak bandwidth \modification{achievable by a server}.
\addition{The datacenter does not contain burst buffers or a distributed file system.}
The peak \addition{bandwidth} of 1.8\,GB/s, obtained from benchmarks run in the datacenter, fits 
high-speed NVMe setups, or RAID-0 over multiple disks or SSDs. %

The \textit{GPU Power Usage}, \textit{GPU temperature}, and \textit{GPU fan Speed} serve as proxy-metrics for GPU load, for which there is no direct utilization metric. %
The GPU power usage is normalized towards the Thermal Design Point (TDP) of each GPU according to Nvidia's official documentation.%
The temperature is normalized against the limits of the GTX 1080ti, Titan V, and RTX Titan which all share the thermal threshold of 91 degree Celsius according to Nvidia's official documentation.
The \textit{GPU memory usage} depicts how much of the GPU memory is being consumed across the datacenter.
The memory limits for the GPU models are 11GB (GTX 1080ti), 12GB (Titan V), and 24GB (RTX Titan).

\textbf{Observations:} From Figure~\ref{surfing:fig:characterizations:general-resource-usage}, we gain several interesting insights that would not have been possible only with high-level performance metrics.
We observe that the number of jobs incoming does not always overlap with any other metric~(\refmainfinding{surfing:mf:queued}).
Intuitively, one would assume that the load would increase based on an increased number of incoming jobs, but as can be observed and further discussed in Section~\ref{surfing:ssct:characterizations:diurnal-load-analysis}, one or more nodes \modification{peak continuously to high levels}---the average system load is typically moderate~(18.2\% of the measurements), high (44.6\%), or very high~(20.2\%)~(\refmainfinding{surfing:mf:lisa-high-load}).
We also observe that the power consumption reaches high levels most of the time.
This suggest that, combined with the observed CPU load, that there is little to no room to deploy energy saving techniques such as dynamic voltage scaling.
Next, we observe that some resources are barely used to their full potential, most notable RAM~(\refmainfinding{surfing:mf:lisa-low-ram-usage}) and GPU memory. 
Overall, disk usage is high~(\refmainfinding{surfing:mf:lisa-lots-of-disk-io}).
Interestingly, we do observe that, based on the GPU metrics, whilst there are periods with moderate to heavy load on the GPUs, the load on the GPUs is mainly low~(\refmainfinding{surfing:mf:lisa-gpu-usage-low}). %
Server temperatures are moderate, even with a high rack power consumption, whereas GPU temperatures are moderate to high most of the time.
We also observe periods with heavy disk I/O and sub-moderate CPU load, indicating the system is not used to its full potential; pipelining approaches or other parallel methods could help.

\subsection{Memory and Network}
\label{surfing:ssct:node-ram-usage}

\begin{figure}[t] %
    \centering
    \includegraphics[max width=\linewidth]{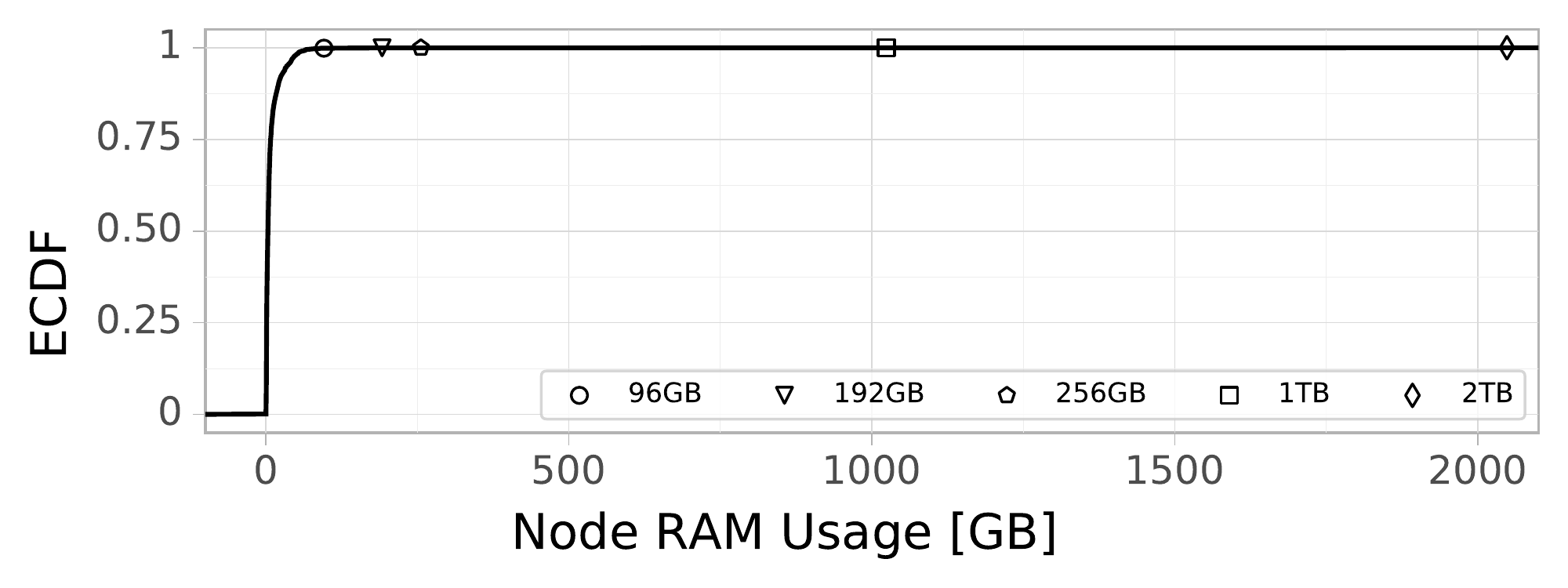}
    \vcutL
    \caption{ECDF of used node RAM. The markers show the different RAM node models in Lisa.}
    \label{surfing:fig:characterizations:node-ram-usage-cdf}
    \vcutM
\end{figure}

\begin{table}[t] %
\caption{RAM usage in the datacenter.}
\label{surfing:tbl:ram-usage-percentiles}
\vcutM
\adjustbox{max width=\linewidth}{
\begin{tabular}{@{}lrrrrrrr@{}}
\toprule
Percentile    & 1\% & 25\% & 50\% & 75\% & 90\% & 99\% & 100\% \\ \midrule
RAM [GB] & 0.64    & 1.46     & 3.65     & 8.07     & 20.99     & 58.06    & 2,000 \\ \bottomrule
\end{tabular}
}
\vcutL
\end{table}

\begin{description}
	\mainfinding{surfing:mf:99-ram-samples}{99\% (75\%) of RAM measurements fit within 64\,GB (8\,GB).}
	\mainfinding{surfing:mf:long-tail-ram}{RAM usage has a very long tail, going up to 2\,TB.}
		\mainfinding{surfing:mf:no-clear-trend-network-job-length}{Longer jobs transmit more network packets, albeit not proportionally with the job duration.} %
	\mainfinding{surfing:mf:outliers}{The longer the job duration, the higher the probability of high outliers for the number of transmitted packets.} %
\end{description}

We characterize RAM usage for the entire dataset;  Table~\ref{surfing:tbl:ram-usage-percentiles} summarizes the basic statistics.
From this table, we observe the RAM usage is low to moderate~(\refmainfinding{surfing:mf:99-ram-samples}). Over 99\% of all RAM measurements are below 59\,GB of RAM, significantly less than the lowest RAM model (96\,GB) in the datacenter. Three-quarters of all RAM measurements fit in 8 GB. %
Combined with the overview in Section~\ref{surfing:ssct:general-resource-usage}, this shows that when designing a datacenter, only few nodes with a lot of RAM are required, reducing costs significantly and being more power efficient.
This is further underlined by the long-tail of RAM usage, with a maximum of 2\,TB~(\refmainfinding{surfing:mf:long-tail-ram}), see Figure~\ref{surfing:fig:characterizations:node-ram-usage-cdf}.

\begin{figure}[t]
    \centering
    \includegraphics[max width=\linewidth]{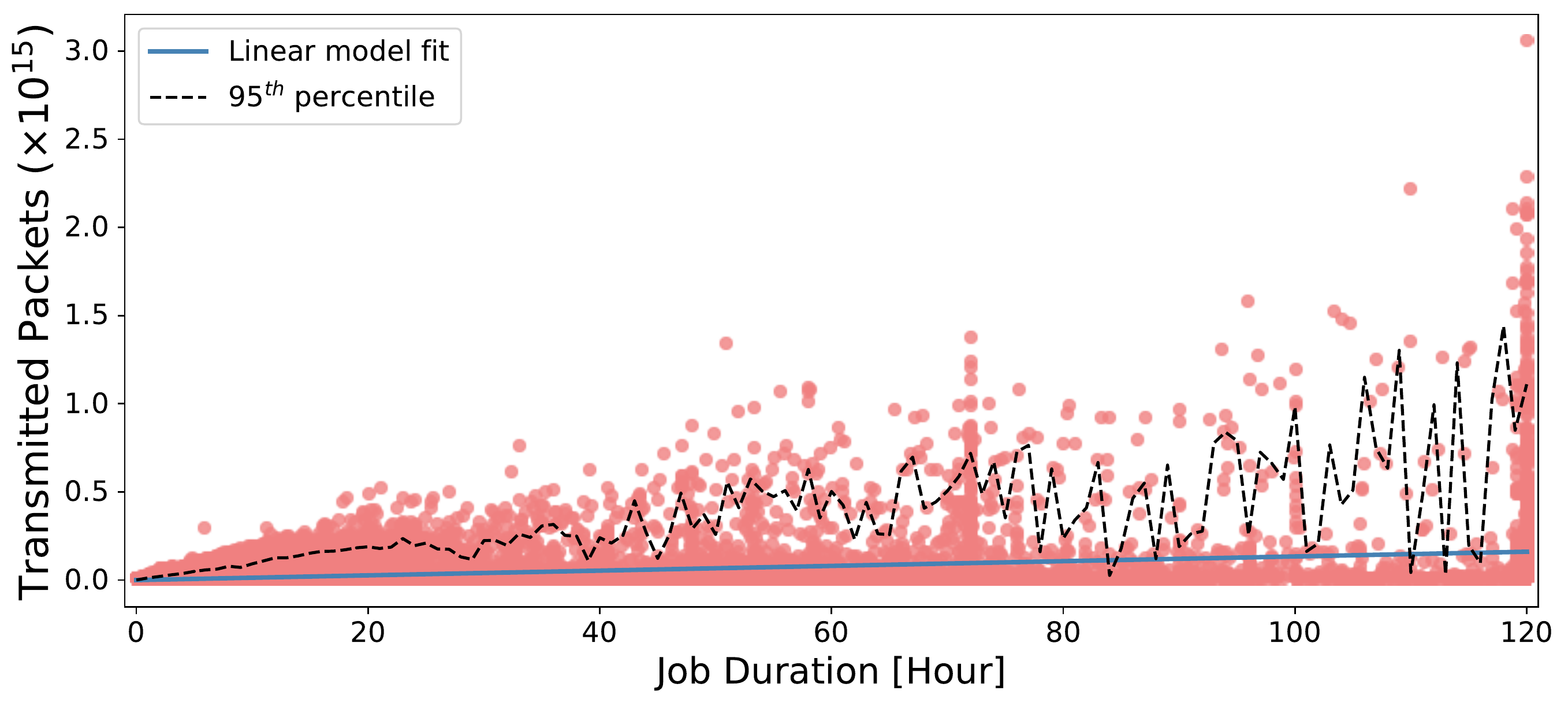}
    \vcutL
    \vcutM
    \caption{Transmitted packets versus job duration.}
    \label{surfing:fig:characterizations:network-analysis:transmitted-packets-job-time-scatterplot}
    \vcutM
    \vcutS
\end{figure}

In Figure~\ref{surfing:fig:characterizations:network-analysis:transmitted-packets-job-time-scatterplot} we plot the number of transmitted packets versus the job time. We observe that shorter jobs seldom send more packets than longer running jobs, i.e., there are no extreme network-heavy yet short-running jobs~(\refmainfinding{surfing:mf:no-clear-trend-network-job-length}).
This could indicate that the majority of the network traffic is in the initial setup, e.g., downloading data. Furthermore, both the number of transmitted packets and the outliers generally increase over time, but only marginally. %
Outliers appear more likely for long-running jobs~(\refmainfinding{surfing:mf:outliers}).
We plot in blue the curve found by the linear regression model fit, which shows that the increase in number of packets transmitted vs job duration is minimal. This could be due to, e.g., MPI jobs generating TCP traffic.
Further analysis that includes more sophisticated network models, e.g., traffic-congestion analysis, is outside the scope of this work but would be possible because the dataset also includes metrics such as TCP retransmission~\cite{DBLP:journals/usenix-login/UtaLIMPC20}.

\subsection{Power Consumption}
\label{surfing:ssct:power-consumption-analysis}

\begin{description}
	\mainfinding{surfing:mf:generic-racks-more-stable}{Generic nodes (racks) have more stable power consumption than ML nodes (racks).} %
	\mainfinding{surfing:mf:consumption}{Generic nodes consume 143\,W on average, and up to 1,300\,W. ML nodes consume 467\,W, and up to $\approx$1,500\,W.}
	\mainfinding{surfing:mf:gpu-and-cpu-racks-exceed-threshold}{Most racks, both generic and ML, exceed the threshold of the cooling system from time to time.}
\end{description}

\begin{figure}[t] %
    \centering
    \includegraphics[max width=\linewidth]{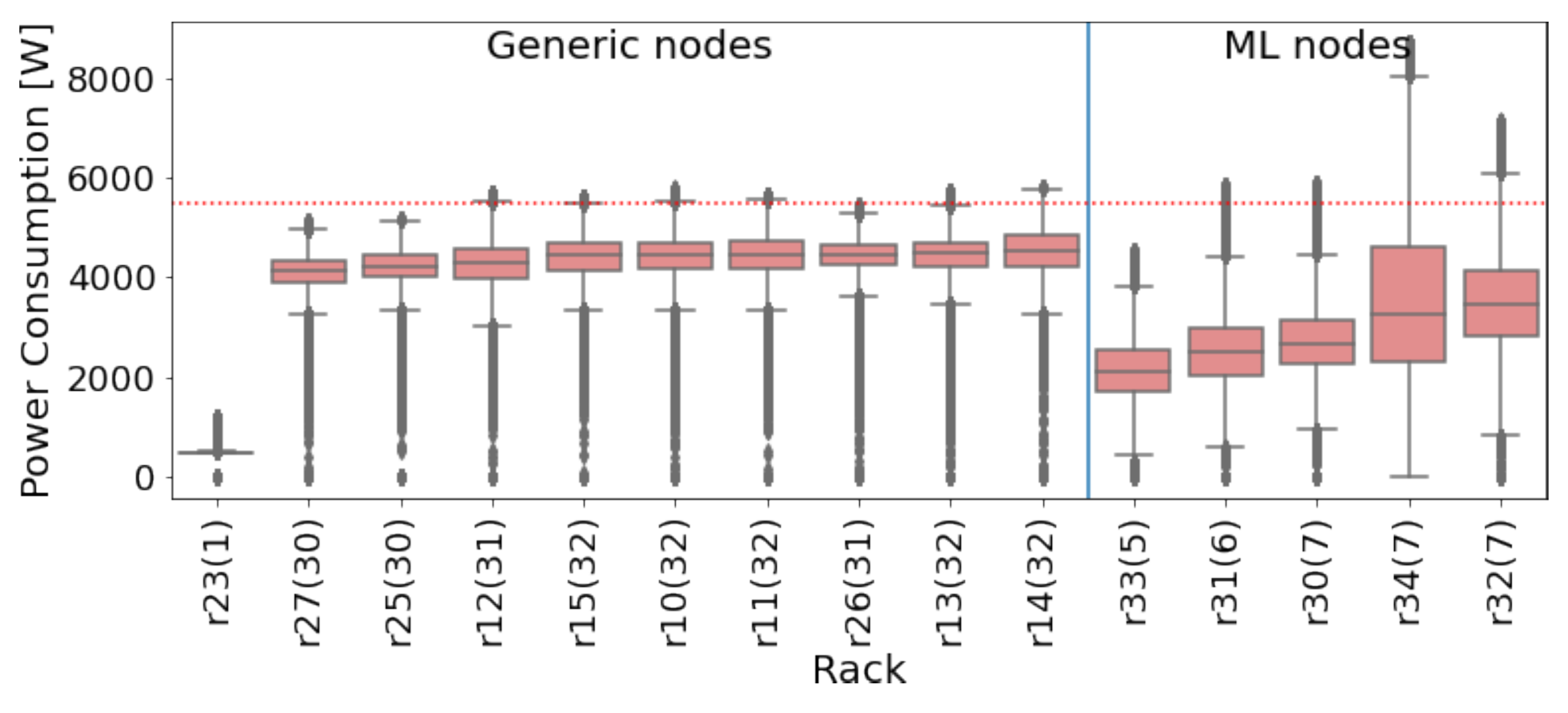}
    \vcutL
    \vcutM
    \caption{Distributions of \textit{rack} power consumption grouped by generic nodes and ML nodes. The labels show between parenthesis how many nodes each rack contains. The distributions are sorted by median per group. The dotted red line depicts the limit of the rack cooling system.}
    \label{surfing:fig:characterizations:power-consumption-per-rack}
    \vcutM
\end{figure}

 \begin{figure}[t] %
     \centering
     \includegraphics[max width=\linewidth]{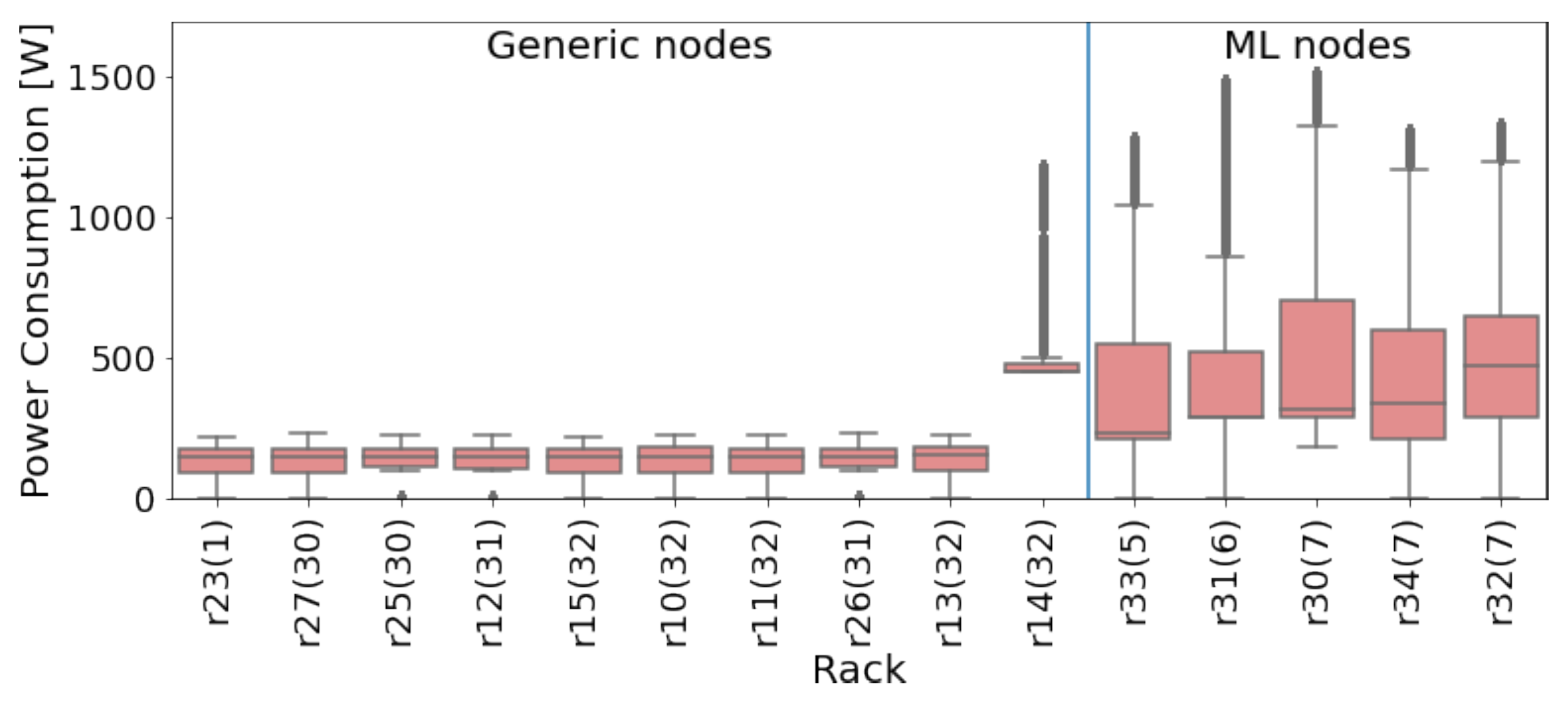}
     \vcutL
     \caption{Distributions of \textit{node} power consumption per rack grouped by generic nodes and ML nodes. The labels show between parenthesis how many nodes each rack contains. The distributions are sorted by median per category.}
     \label{surfing:fig:characterizations:power-consumption-per-node-per-rack}
     \vcutM
 \end{figure}

Energy consumption is becoming increasingly important~\cite{duy2010performance}. To better understand the power consumption within the datacenter, we observe power consumption using two different levels.
First we show the distribution of power consumption per \textit{rack}, in Figure~\ref{surfing:fig:characterizations:power-consumption-per-rack}.
We additionally group together generic nodes and ML nodes as the latter contain accelerators (GPUs).

\begin{table}[t]
\caption{Power consumption (Watt) of generic and ML \textit{nodes}.}
\label{surfing:tbl:power-consumption-percentiles-per-node-type}
\vcutM
\adjustbox{max width=\linewidth}{
\begin{tabular}{@{}lrrrrrrrr@{}}
	\toprule
	& 1\% & 25\% & 50\% & mean & 75\% & 99\% & 100\% \\ \midrule
	Generic & 80.00    & 100.00     & 148.00     & 143.01     & 176.00     & 260.00 & 1,300.00   \\
	ML      & 130.00    & 260.00     & 364.00     & 467.16     & 624.00     & 1,274.00 & 1,508.00    \\ \bottomrule
\end{tabular}
}
\vcutL
\end{table}

Figure~\ref{surfing:fig:characterizations:power-consumption-per-rack} shows that there is little to moderate variation in generic node rack power consumption, with the exception of rack 23.
Furthermore, the IQR ranges of the boxplots within each violin plot show that most generic racks consume more power compared to ML racks.
The ML racks show more variation and have higher extremes even though they contain fewer nodes~(\refmainfinding{surfing:mf:generic-racks-more-stable}), see Table~\ref{surfing:tbl:power-consumption-percentiles-per-node-type}.
The fluctuations are due to the power profile of GPUs: idle they consume as little as 1 Watt, yet at full load their power consumption goes up as high as 416 Watts.
As ML nodes have up to four GPUs, the power consumption can go significantly higher than generic nodes.
The reason for rack 23 being an outlier is that it only hosts one node vs. 30-32 for the other generic node racks.
Hence, this causes a lower power consumption profile for the rack.

Next, we investigate the power consumption of individual nodes within each rack.
From Figure~\ref{surfing:fig:characterizations:power-consumption-per-node-per-rack} and 
Table~\ref{surfing:tbl:power-consumption-percentiles-per-node-type} we observe that generic nodes feature a small range, typically between 80-260 Watts. 
One exception is rack 23 whose distribution is more than 3x higher compared to the other generic nodes. This is the only node with four sockets where its CPUs have a higher TDP than most of the other nodes.
The node contains 48 CPU cores, 4x more than the regular generic nodes, in line with the 3x increase in power consumption(~\refmainfinding{surfing:mf:consumption}). 
Moreover, the node contains significantly more RAM (see also Section~\ref{surfing:ssct:node-ram-usage}) which means additional power draw.

Comparing the generic nodes with the ML nodes, we observe the generic nodes power consumption range is constrained, which in turn limit the ranges of the racks.
As the generic node racks pack more nodes, they consume more energy, leading to the higher average seen in the previous discussed image.
We also wondered if the lower number of nodes per ML rack is due to power supply unit or cooling system limitations.
After inquiring the datacenter operators, the cooling system is indeed the limiting factor, only handling loads up to 5.5\,kW per rack.
We observe these are occasionally exceeded (\refmainfinding{surfing:mf:gpu-and-cpu-racks-exceed-threshold}). Datacenter designs that include accelerators or aim for upgradeability have to consider this power-limiting aspect, underlined by the recently announced GPUs by Nvidia whose power consumption increased significantly\footnote{see e.g., \url{https://www.tomshardware.com/news/nvidias-rtx-3000-power-supply-requirements-PSU-shortage-2020}} compared to older versions.

\subsection{Rack Temperature}
\label{surfing:ssct:characterizations:temp-analysis}

\begin{figure}[t]
    \centering
    \begin{subfigure}{(a) 7 distinct nodes in an ML rack (rack number 32).}
    \vcutM
        \includegraphics[width=0.99\columnwidth]{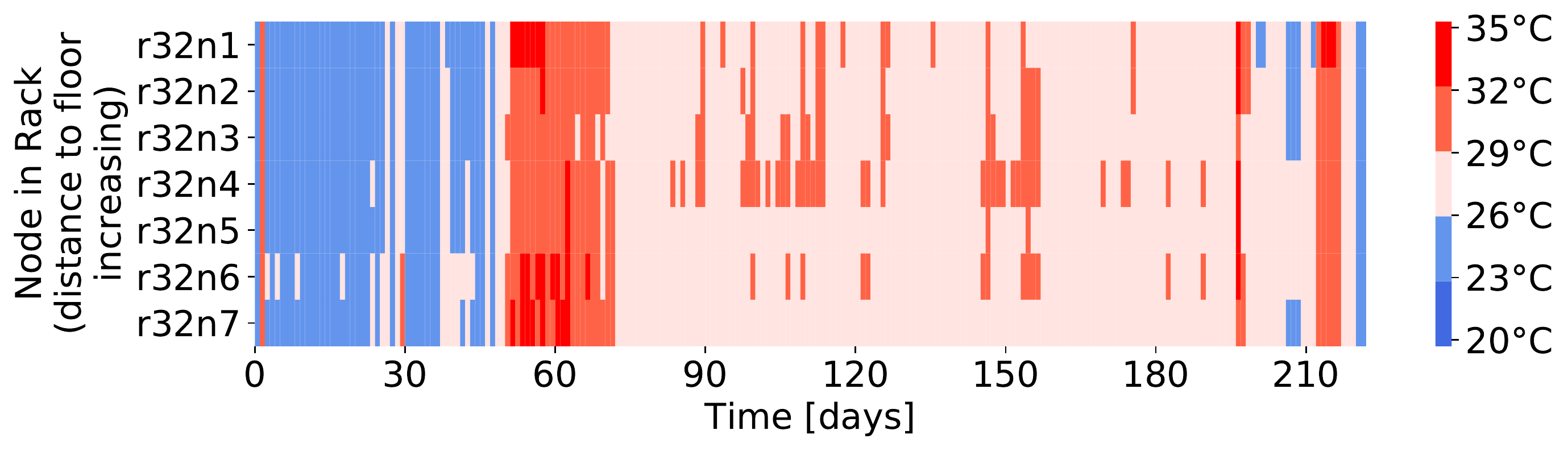}
    \end{subfigure}
    \begin{subfigure}{(b) 32 distinct nodes in a generic rack (rack number 10).}
    \vcutM
        \includegraphics[width=0.99\columnwidth]{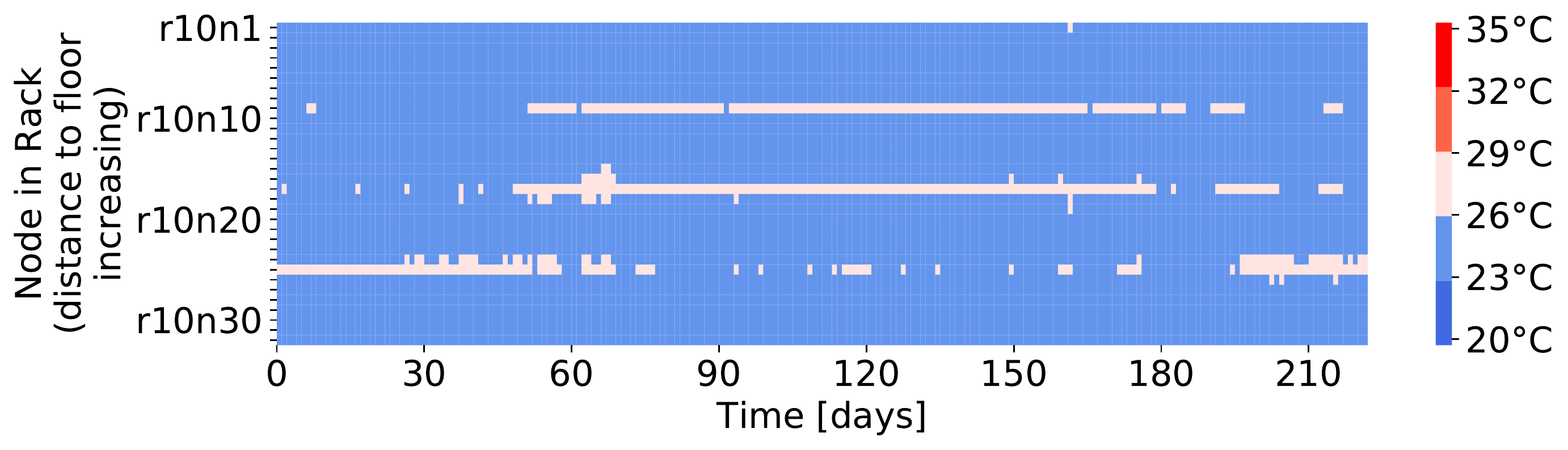}
    \end{subfigure}
    \vcutL
    \caption{Max daily temperature, ML vs. generic nodes.} %
    \label{surfing:fig:characterizations:temp-analysis:gpu-rack}
    \vcutM
\end{figure}

\begin{description}
\mainfinding{surfing:ssct:characterizations:temp-analysis1}{Temperature is correlated between nodes in the same rack.}%
\mainfinding{surfing:ssct:characterizations:temp-analysis2}{Temperature and node position in rack are not correlated.}
\mainfinding{surfing:ssct:characterizations:temp-analysis3}{GPU-racks run hotter than CPU-only racks, by $\approx$3\textdegree{}C.} %
\end{description}

The dataset we analyze in this paper contains multiple types of temperature-related metrics: GPU temperature, as well as server ambient temperature. 
While the former is the chip temperature, which is highly correlated with GPU workload, the latter is the temperature inside the server enclosure, which is influenced by many other factors: CPU workload, cooler (mal)functioning, as well as warmer nearby nodes and distance from the datacenter floor.
According to the datacenter operator, all nodes in this study are air cooled.

We find that nodes in ML racks tend to be correlated in terms of temperature (\refmainfinding{surfing:ssct:characterizations:temp-analysis1}).
They are either mostly warmer, or mostly cooler. Figure~\ref{surfing:fig:characterizations:temp-analysis:gpu-rack} plots this behavior. 
The graph shows the maximum temperature registered by servers in rack 32 for the entire period the dataset was collected. 
The graph also maintains the server ordering in the rack, with the smallest node id at the top (see vertical axis). We notice that the node positioning in the rack does not influence its temperature (\refmainfinding{surfing:ssct:characterizations:temp-analysis2}).
\addition{This finding matches the type of cooling used, i.e., air. Based on the experience of the datacenter operator, water cooling would not change the conclusion, because water cooling has superior heat dissipation.}
For the entire period, the lowest node temperature is around 20\textdegree{}\,C, while the highest temperature is 35\textdegree{}\,C. 
This range is significantly lower than reported by Netti et al.~\cite{DBLP:conf/hpdc/NettiMGOTO020} where a range of 47-54\textdegree{}\,C is reported.
The difference could be caused by a different node hardware and cooling system combination.
The figure depicts clearly that hotter periods are correlated over the entire rack. This type of behavior holds for all ML racks.
\addition{If water cooling is used, it's likely that these temperatures would remain low and thus will not correlate as observed, due to the efficiency of these systems~\cite{zhang2018comparison}.}
The generic racks are much cooler: most nodes operate at 23-25\textdegree{}\,C, $\approx$3\textdegree{}\,C lower than most ML-rack nodes (\refmainfinding{surfing:ssct:characterizations:temp-analysis3}).

\subsection{CPU Diurnal Load}
\label{surfing:ssct:characterizations:diurnal-load-analysis}

\begin{figure}[t]  %
	\centering
	\includegraphics[max width=\linewidth]{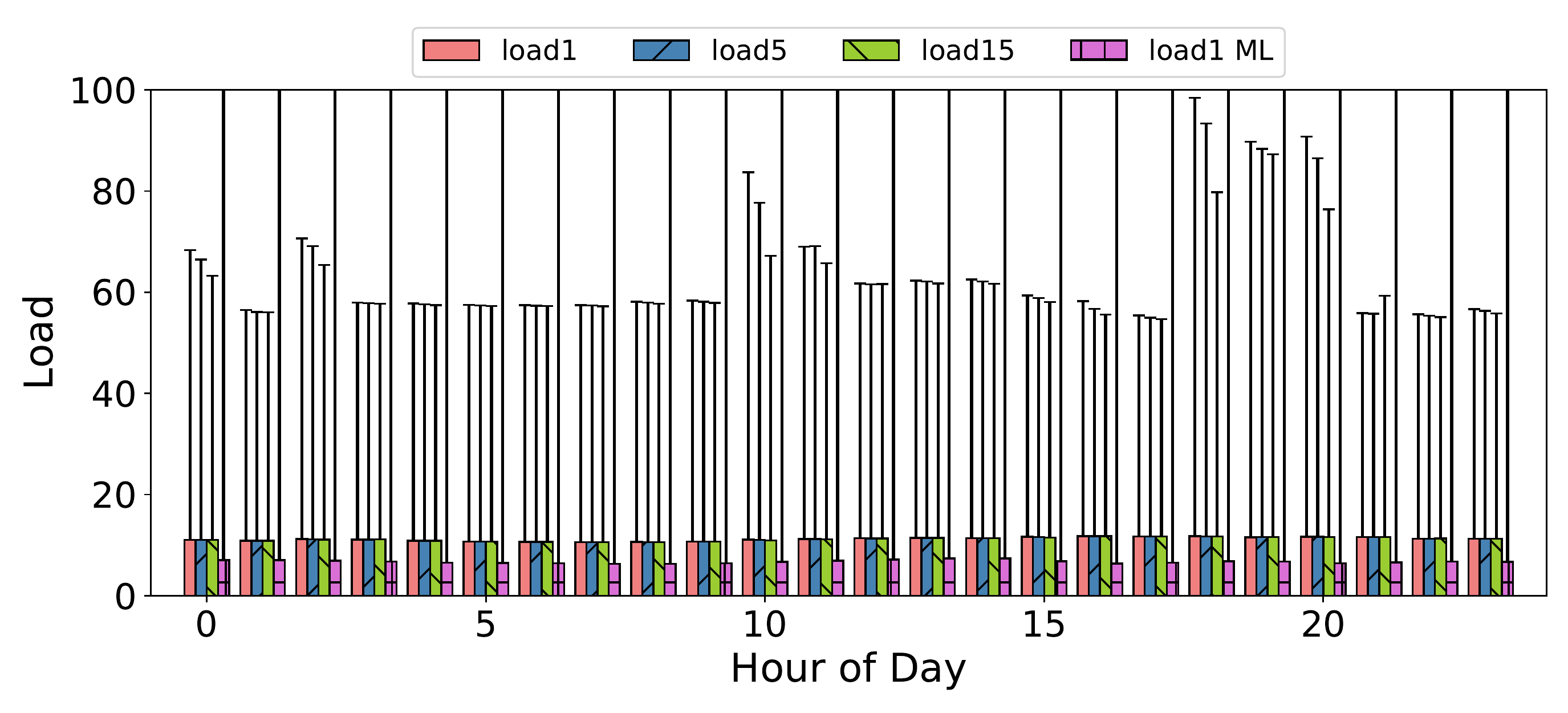}
	\vcutL
	\vcutM
	\caption{The average UNIX load1, load5, and load15 metrics per hour of day. The error bars depict the standard deviation.}
	\label{surfing:fig:characterizations:diurnal-load-analysis:diurnal-hourly}
	\vcutL
\end{figure}

\begin{figure}[t]  %
	\centering
	\includegraphics[max width=\linewidth]{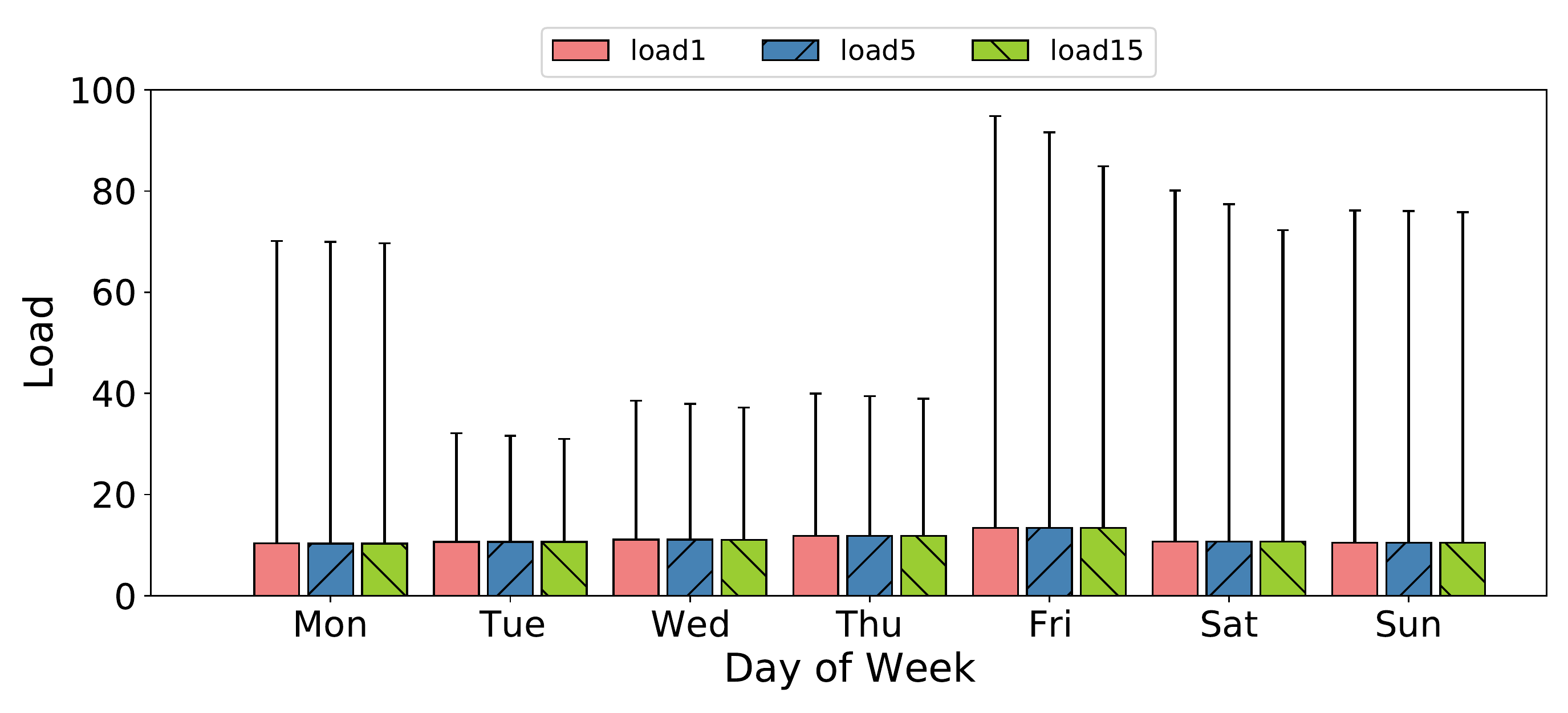}
	\vcutL
	\vcutM
	\caption{The average UNIX load1, load5, and load15 metrics per day of week. The error bars depict the standard deviation.}
	\label{surfing:fig:characterizations:diurnal-load-analysis:daily}
	\vcutL
\end{figure}

\begin{description}
    \mainfinding{surfing:ssct:characterizations:diurnal1}{The average system loads are stable. (See also \refmainfinding{surfing:mf:job-submissions-lisa-diurnal-pattern}.)}
    \mainfinding{surfing:ssct:characterizations:diurnal2}{Across all hours, ML nodes have an average load1 metric $\approx$40\% lower than generic nodes.}
\end{description}

To investigate the daily and weekly trends that may appear in the datacenter, we depict in Figure~\ref{surfing:fig:characterizations:diurnal-load-analysis:diurnal-hourly} the load1, load5, and load15 UNIX metrics across the entire datacenter. We notice that the average load is very stable within the datacenter (\refmainfinding{surfing:ssct:characterizations:diurnal1}).
The averages range between 10.6 and 11.8 for load1.
 Interestingly, this does not match the arrival pattern of jobs visible in Figure~\ref{surfing:fig:num-jobs-hour-of-day}.
This might be due to the loads being regularly above 16, depicted by the error bars. This behavior indicates that processes are getting delayed as the most common node within the datacenter features 16 cores. 

Figure~\ref{surfing:fig:characterizations:diurnal-load-analysis:daily} presents the load per day of week.
We observe that the load in the cluster is also stable through the weeks, which aligns with the average per hour of day.
There is a minimal elevation on Fridays and a small decrease in the weekend. 
Similarly to hour of day, the arrival of jobs does not correlate with the load.

When considering the load1 of ML nodes, we notice that it is stable, yet significantly lower than the cluster average.
The average load1 per hour ranges between 6.3 and 7.4, which is around 40\% lower than the average load across all machines (\refmainfinding{surfing:ssct:characterizations:diurnal2}).
This indicates that these machines are utilized less.
In Section~\ref{surfing:sct:workload-characterization}, where we characterize the workload in-depth, we notice in Figure~\ref{surfing:fig:cov-cpus-requested-users} that indeed fewer users submit ML jobs.

%% file: 05_covid.tex
\section{Operations in the Time of COVID-19} %
\label{surfing:sct:covid}
\label{surfing:sec:covid}

\begin{figure}[t]
    \centering
    \adjustbox{width=\linewidth}{
    \includegraphics[max width=\linewidth]{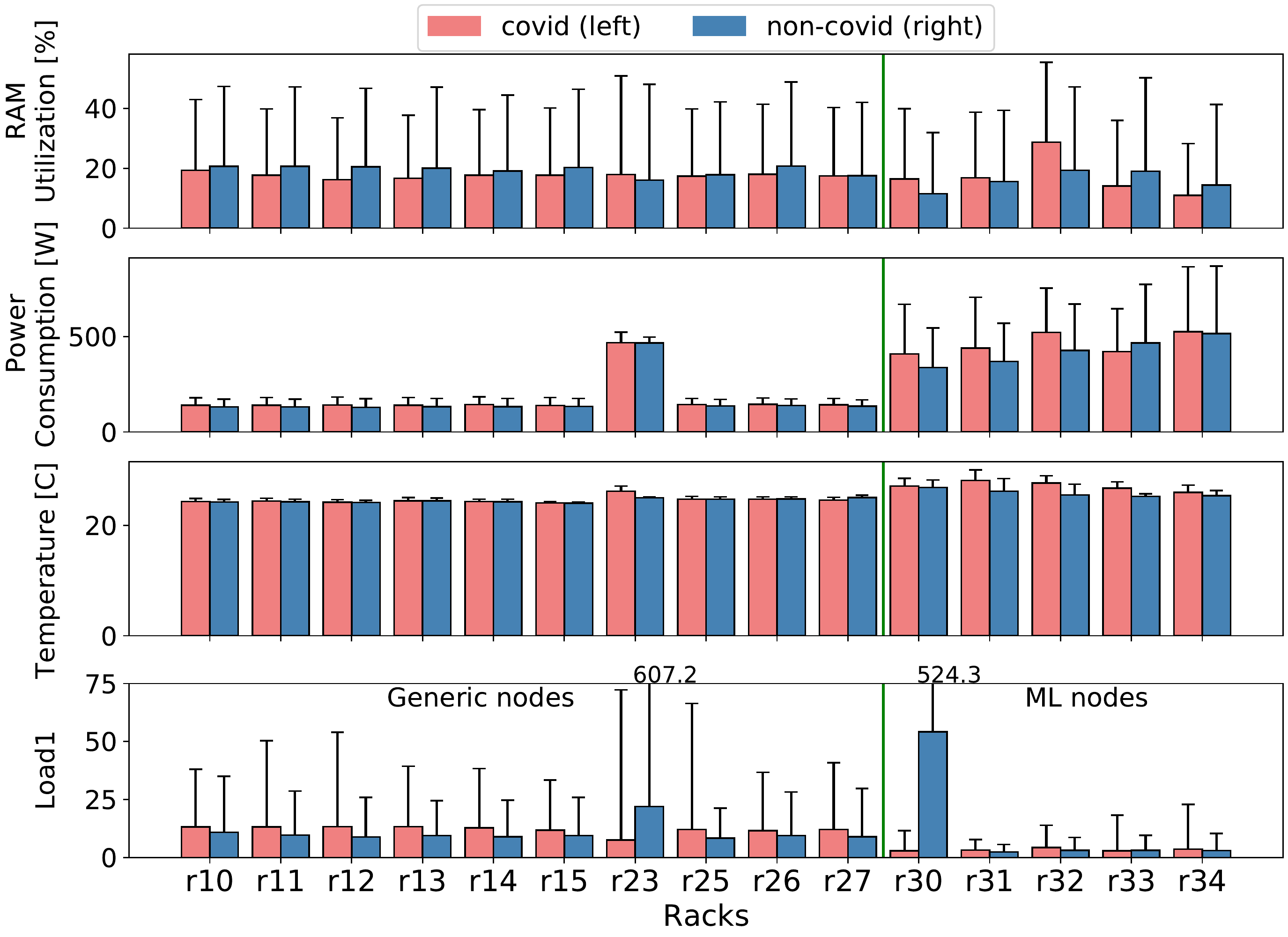}}
	\vcutL
	\vcutM
    \caption{The average RAM utilization, power consumption, ambient temperature, and load1 values for each node, aggregated per rack. The error bars depict the standard deviation.}
    \label{surfing:fig:covid-rack-barplot}
    \vcutL
\end{figure}

We analyze how the datacenter operations changed during the 2020 COVID-19 pandemic, with the method from Section~\ref{surfing:sec:method:analysis}.
We analyze data per rack and per note.
Figure~\ref{surfing:fig:covid-rack-barplot} depicts the most important results. %
The results in this section lead us to one main observation:
\begin{description}
	\mainfinding{surfing:mf:covidoverall}{During the covid period, average resource utilization, power consumption, and temperature did not increase significantly. Covid operations continued without large discrepancies.}
\end{description}
This finding was surprising to the datacenter operation team, and could have positive implications for workload procurement.

\begin{figure}[t]
	\centering
	\adjustbox{width=\linewidth}{
		\includegraphics[max width=\linewidth]{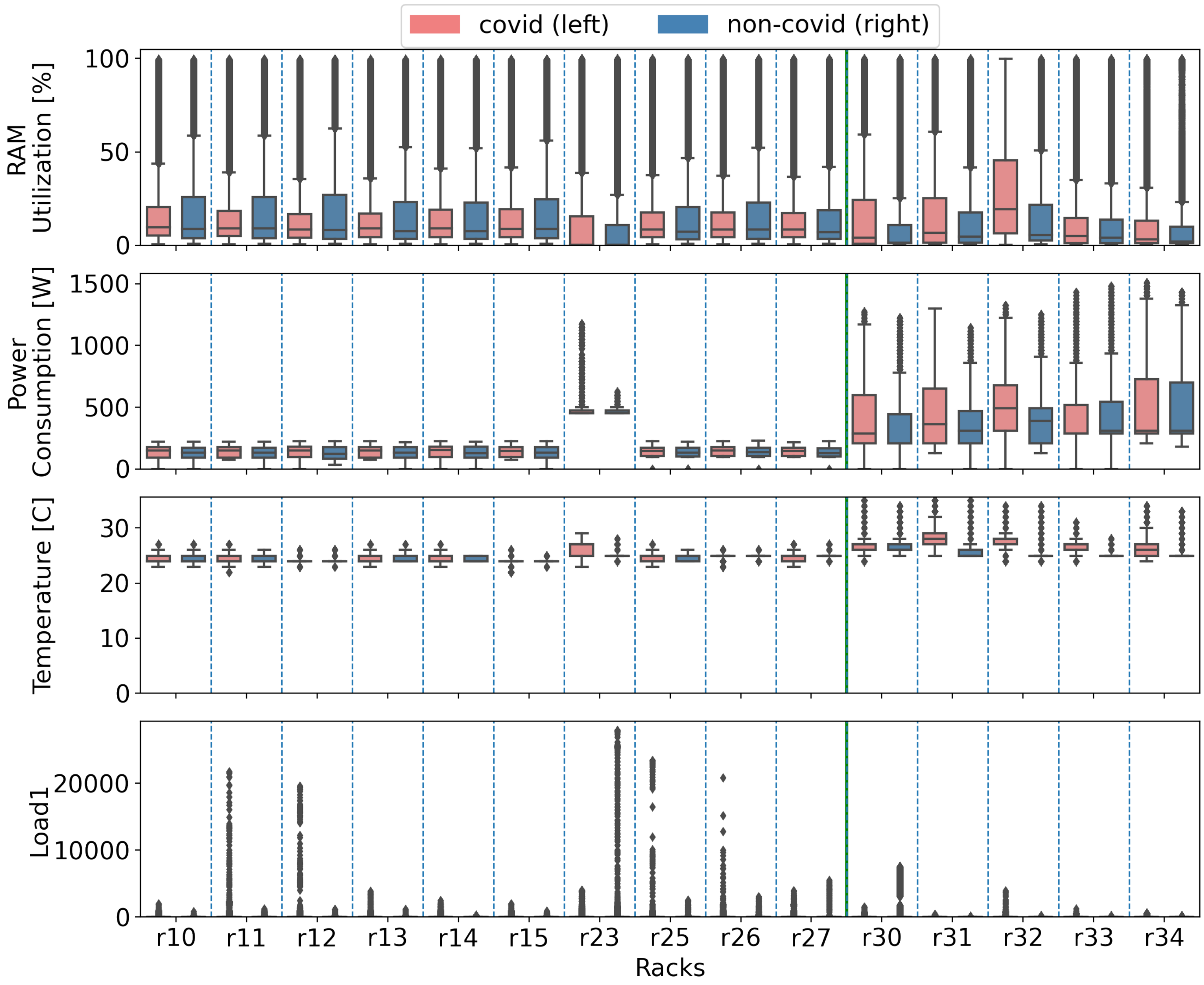}}
	\vcutL
	\caption{RAM utilization, power consumption, ambient temperature, and load1 value distribution for each node, per rack. The left side of the green vertical line are the generic racks.}
	\label{surfing:fig:covid-rack-boxplots}
	\vcutM
\end{figure}

\begin{figure}[t]
    \centering
    \adjustbox{width=\linewidth}{
    \includegraphics[max width=\linewidth]{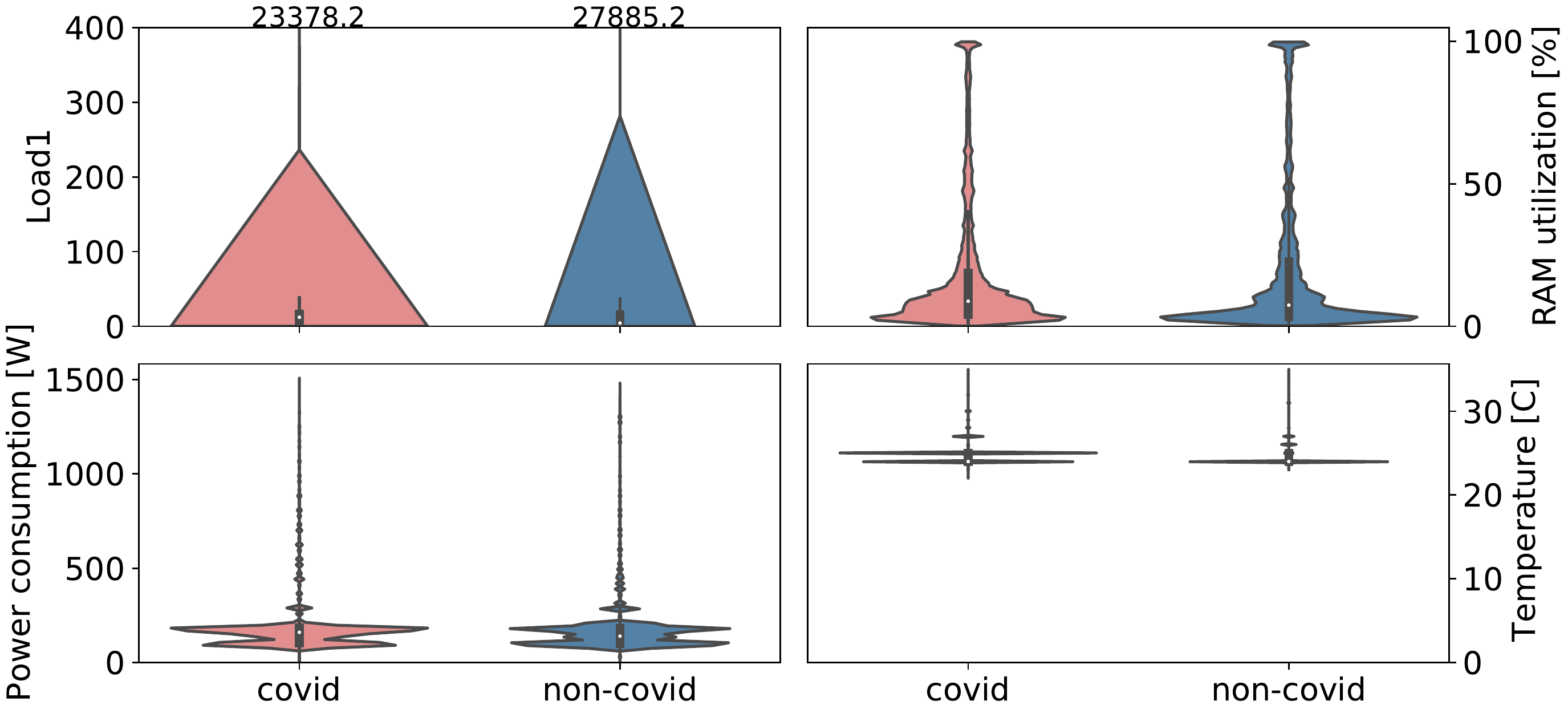}}
	\vcutL
    \caption{Load1, power consumption, RAM utilization and ambient temperature value distribution for the covid and non-covid periods across all the nodes.}
    \label{surfing:fig:covid-cluster-violinplot}
    \vcutM
\end{figure}

\subsection{Memory Usage}   \label{surfing:ssct:covid-memory-analysis}
\begin{description}
	\mainfinding{surfing:mf:covid-lower-memory-usage-generic-nodes-covid}{During covid, average RAM usage generally decreases in generic racks, but not in ML racks. %
	}
\end{description}

The average RAM utilization is 1\% to 5\% higher during the non-covid period than during the covid period, for the nine generic racks~(\refmainfinding{surfing:mf:covid-lower-memory-usage-generic-nodes-covid}); only one rack exhibits higher RAM utilization during the covid period. 
This can also be observed in Figure~\ref{surfing:fig:covid-rack-boxplots} where both the IQRs and the whiskers are higher for the respective boxplots.
From the ML racks, changes in RAM usage are mixed: 3 racks exhibit a decrease, 2 racks an increase.
We conjecture the specialist use of ML racks makes it less likely to change behavior; in the Netherlands, experts continued work without much disruption, remotely.

Across all nodes, the RAM utilization is slightly higher in the non-covid period, see Figure~\ref{surfing:fig:covid-cluster-violinplot}.

\subsection{Power Consumption and Temperature}  \label{surfing:ssct:covid-power-consumption-analysis}
\label{surfing:ssct:covid-temperature-analysis}
    
\begin{description}
	\mainfinding{surfing:mf:power-no-difference-covid-generic}{No difference in power consumption for generic nodes.}
	\mainfinding{surfing:mf:power-no-difference-covid-ml}{Moderately increased power consumption, for ML.}
	\mainfinding{surfing:mf:temperature-no-difference-covid-generic}{No to low temperature increase for generic nodes.}
	\mainfinding{surfing:mf:temperature-ml}{Low temperature increase for ML nodes.}
\end{description}

Figure~\ref{surfing:fig:covid-rack-barplot} also shows that three ML node racks consume on average 50\,W to 100\,W more power during the covid period~(\refmainfinding{surfing:mf:power-no-difference-covid-ml}). This is a moderate increase, relatively to the values in Section~\ref{surfing:ssct:power-consumption-analysis}. 

Nodes in the generic racks do not exhibit a similar behavior~(\refmainfinding{surfing:mf:power-no-difference-covid-generic}).
The stability of the power consumption for the generic nodes, introduced in Section~\ref{surfing:ssct:power-consumption-analysis}, also appears here, including for the prior outlier rack r23. 
We attribute this phenomenon to the datacenter continuing regular service during covid, but onboarding fewer inexperienced users that could introduce variable load while learning how to use the system.

Interestingly, rack 23 does show more and more extreme outliers in the covid period, see Figure~\ref{surfing:fig:covid-rack-boxplots}.
With a significant RAM utilization (i.e., 50+\%), the node appears to be used more intensely during the covid period. 
For both periods, ML node racks consume more power than the generic node racks (with the exception of special rack 23).

The temperature for both periods is very close to 25\textdegree{}C on average, which is the ideal temperature for servers~\cite{pedram2012energy}. 
The generic nodes do not exhibit temperature increases in general; only rack 23 exhibits a 2-3\textdegree{}C increase~(\refmainfinding{surfing:mf:temperature-no-difference-covid-generic}).
Except for rack r30, the ML node racks have 1\textdegree{}C to 3\textdegree{}C higher temperature during the covid period, especially rack r31~(\refmainfinding{surfing:mf:temperature-ml}).
These elevations also show in the boxplots, see Figure~\ref{surfing:fig:covid-rack-boxplots}.   %
However, these increases of just a few degrees still correspond to normal operation.
We conclude there are no significant temperature differences between the covid and non-covid periods.

\subsection{System Load}   \label{surfing:ssct:covid-load-analysis}
\begin{description}
	\mainfinding{surfing:mf:corona-load-higher-generic-racks}{Increased load for several generic racks, during covid.}
	\mainfinding{surfing:mf:corona-load-same-lower-ml-racks}{ML racks unchanged. Rack 30 decrease during covid.}
\end{description}

The generic node racks, except for rack r23, have a higher average and significantly higher outliers during the covid period.
This can be observed in Figure~\ref{surfing:fig:covid-rack-boxplots}.
This indicates nodes utilized heavier, and particularly using short, heavy bursts, as the average remains low in comparison to the values of the peaks, see Figures~\ref{surfing:fig:covid-rack-barplot} and \ref{surfing:fig:covid-cluster-violinplot}~(\refmainfinding{surfing:mf:corona-load-higher-generic-racks}). The reverse holds for Rack 23, which is surprising giving the previous metrics show an elevation. %
As the node has 48 CPU cores, as outlined in Section~\ref{surfing:ssct:power-consumption-analysis}, 
this node is rarely overloaded.
These results suggest the jobs submitted during the covid period generate fewer tasks to be processed in parallel by the individual cores, yet they do lead to more power consumption and RAM utilization, which in turn could cause the elevation in temperature.

From the ML node racks, only Rack 30 has significantly different load---much lower load during the covid period.
Nodes in the other ML racks exhibit no significant load differences during covid~(\refmainfinding{surfing:mf:corona-load-same-lower-ml-racks}).

%% file: 06_workload_characterization.tex
\section{Workload Characterization}
\label{surfing:sct:workload-characterization}
\label{surfing:sec:workload}

We characterize in this section the datacenter workload, with the method from Section~\ref{surfing:sec:method:analysis}.

\subsection{Job Characteristics}
\label{surfing:ssct:dataset-outline-job-characteristics}

\begin{figure}[t]
	\centering
	\includegraphics[max width=\linewidth]{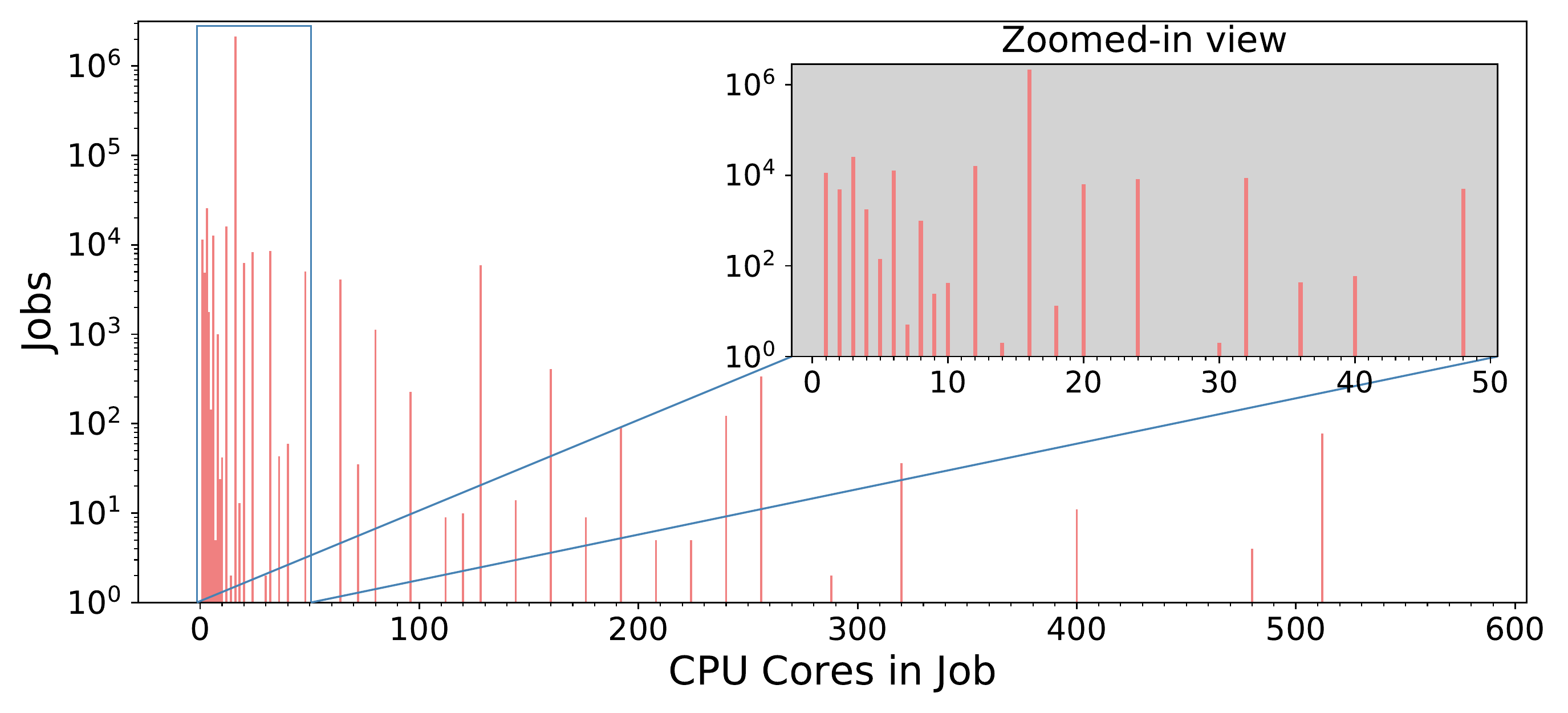}
	\vcutL
	\vcutM
	\caption{Frequency distribution of allocated CPU-cores.}
	\label{surfing:fig:characterizations:job-cpus-count}
	\vcutM
\end{figure}

\begin{description}
    \mainfinding{surfing:mf:lisa-jobs-are-small}{Most jobs are small. Most jobs request less than 100 CPU cores, with a mode of 16 cores and max $>$500.}
    \mainfinding{surfing:mf:lisa-jobs-are-short}{Most jobs are short: $\approx$90\% of all completed jobs have a runtime $\leq$300\,seconds.}
\end{description}

For job sizes, we depict the frequency of allocated CPU-cores in Figure~\ref{surfing:fig:characterizations:job-cpus-count}~(next page).
Most jobs are small~(\refmainfinding{surfing:mf:lisa-jobs-are-small}).
Considering the number of requested cores (equal to the number of allocated cores in this system), Figure~\ref{surfing:fig:characterizations:job-cpus-count} features a peak for 16 cores. This is equal, for example, to the number of requested cores in the Google trace~\cite{DBLP:conf/usenix/AmvrosiadisPGGB18}.
As the most common nodes in the system have 16 cores, we believe most users simply request one full node using SLURM; the largest queue in SLURM enables this behavior. 
Most submitted jobs request less than 100 CPU cores, with extremes using over 500 CPU cores~(\refmainfinding{surfing:mf:lisa-jobs-are-small}). Few users queue large jobs as, depending on the job-placement policy, it can take a considerable amount of time before enough resources become available.

\begin{figure}[t]
	\centering
	\includegraphics[width=0.9\linewidth]{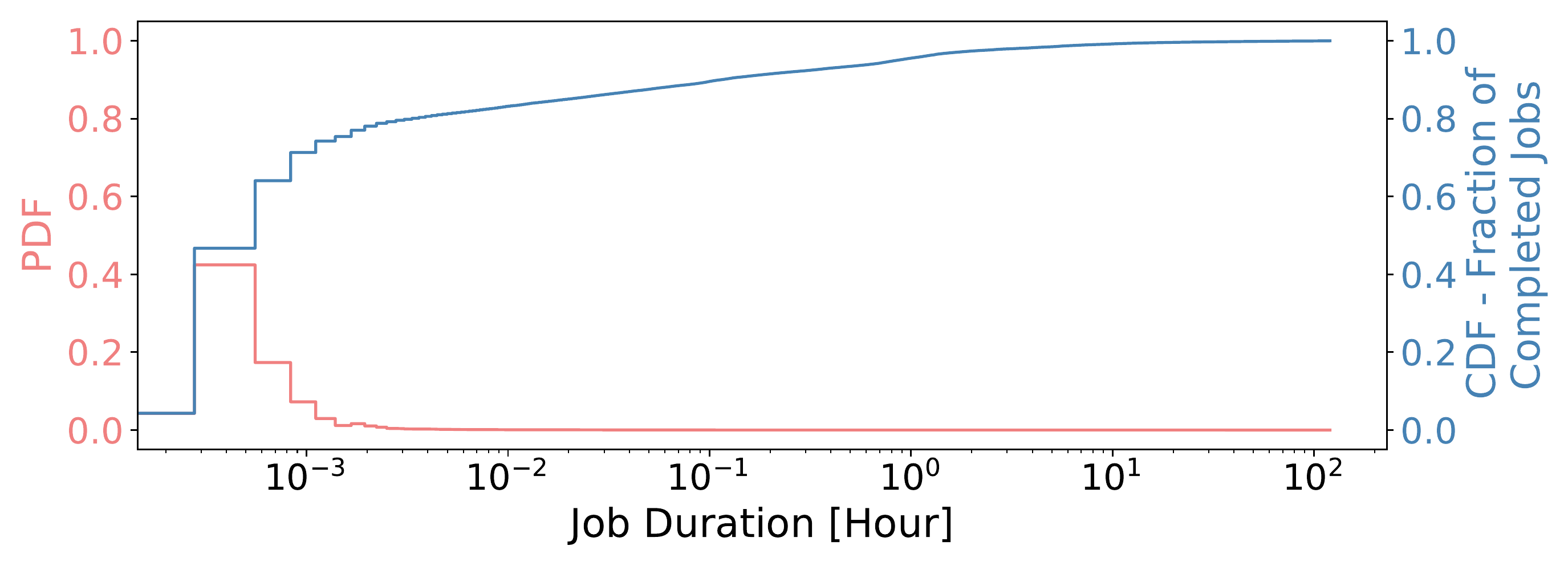} 
	\vcutL
	\caption{Duration of completed jobs, CDF-PDF plot.}
	\label{surfing:fig:characterizations:job-duration-cdf-pdf} 
	
\end{figure}

\begin{figure}[t]
	\centering
	\includegraphics[max width=\linewidth]{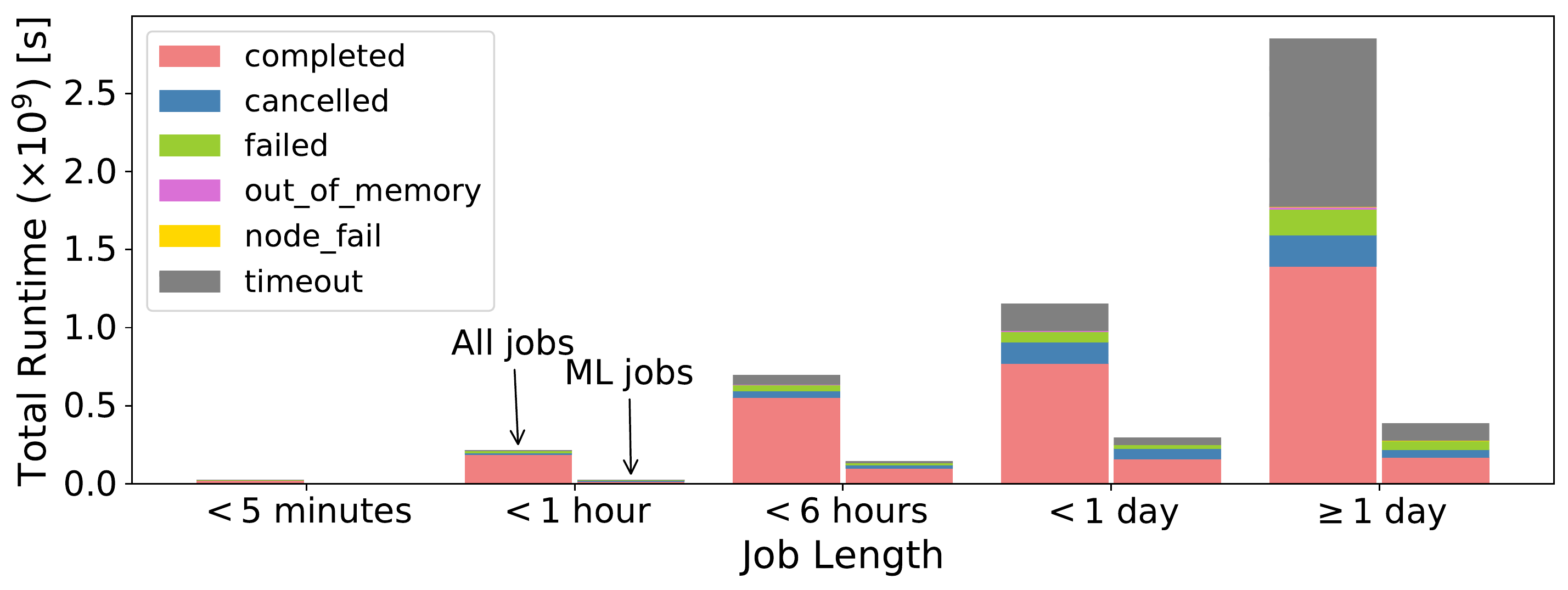}
	\vcutL
	\vcutM
	\caption{Total job runtime grouped by job length. Per bar, we stack the runtimes per job state.}
	\label{surfing:fig:job-outcome-per-runtime-category}
	\vcutM
\end{figure}

\begin{figure}[t]
	\centering
	\includegraphics[max width=\linewidth]{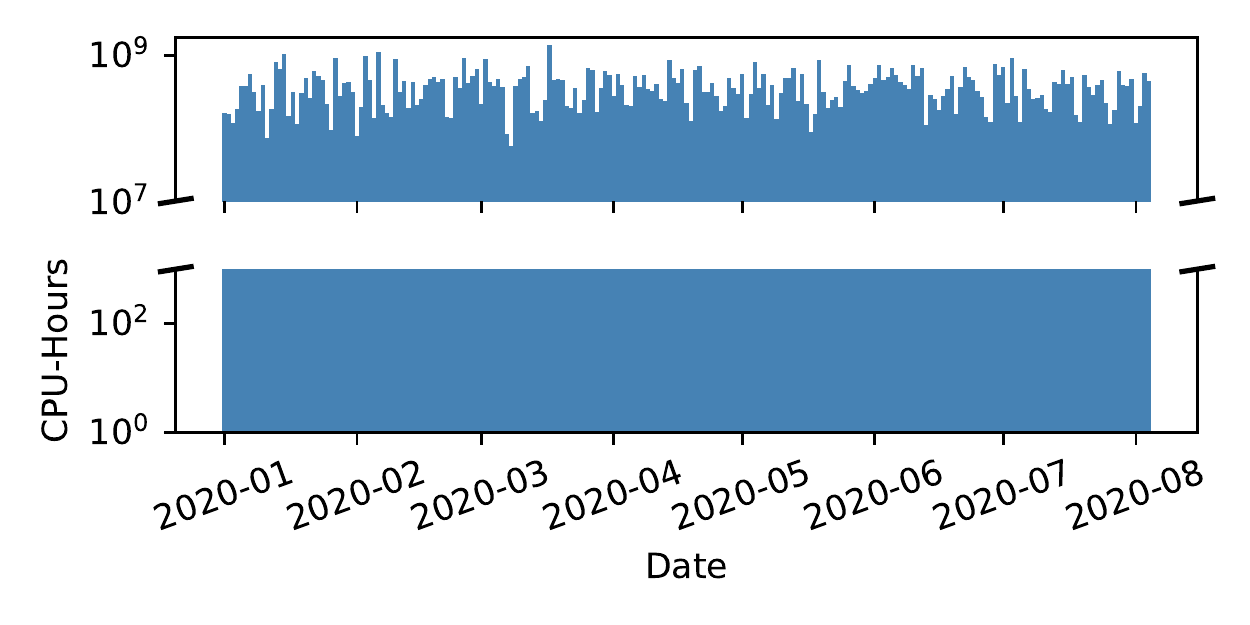}
	\vcutL
	\caption{Overview of the daily 'footprint' of submitted jobs.}
	\label{surfing:fig:characterizations:job-squashed-area}
	\vcutM
\end{figure}

We inspect the runtime of jobs within the datacenter. Figure~\ref{surfing:fig:characterizations:job-duration-cdf-pdf} shows the CDF of job durations. Most jobs are short: 88.9\% of all completed jobs have a runtime of 5\,minutes or less~(\refmainfinding{surfing:mf:lisa-jobs-are-short}). Figure~\ref{surfing:fig:job-outcome-per-runtime-category} shows short-jobs also consume less, cumulatively, than long-running jobs. 
The cumulative runtime of short jobs is more than $177\times$ smaller than for jobs running for a day or longer.
Interestingly, jobs lasting up and until one hour take up a noticeably larger share when compared to other publicly available cluster traces~\cite{DBLP:conf/usenix/AmvrosiadisPGGB18,DBLP:journals/fgcs/IosupLJADWE08,DBLP:conf/ccgrid/ShenBI15}. %

\subsection{Arrival Patterns}
\label{surfing:ssct:dataset-outline-workload-heterogeneity}

\begin{figure}[t]
	\centering
	\includegraphics[max width=\linewidth]{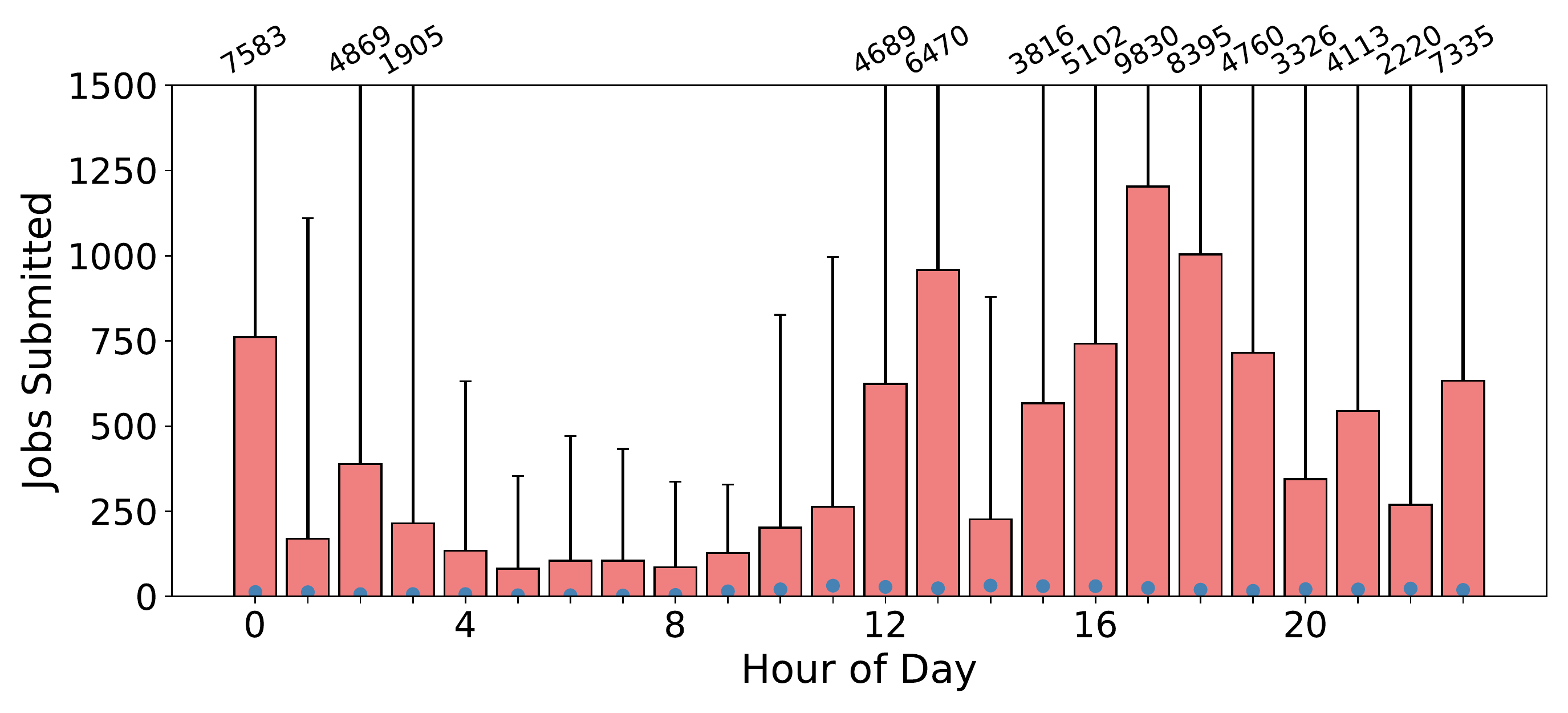}
	\vcutL
	\vcutM
	\caption{Number of submitted jobs per hour of day. The blue dots depict the number of ML jobs.}
	\label{surfing:fig:num-jobs-hour-of-day}
	\vcutM
\end{figure}

\begin{figure}[t]
	\centering
	\includegraphics[max width=\linewidth]{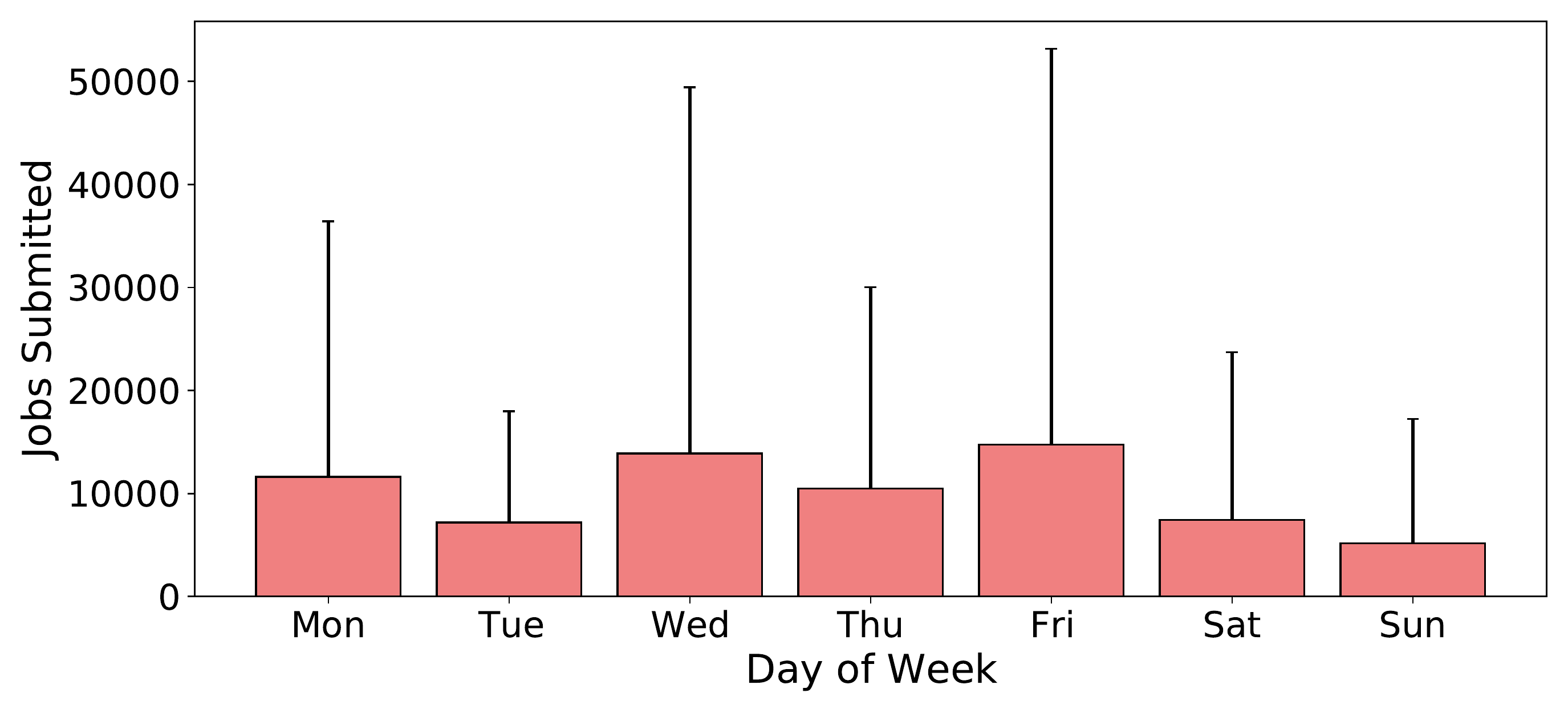}
	\vcutL
	\vcutM
	\caption{Number of submitted jobs day of week.}
	\label{surfing:fig:num-jobs-day-of-week}
	\vcutL
\end{figure}

\begin{figure}[t]
	\centering
	\includegraphics[max
	 width=\linewidth]{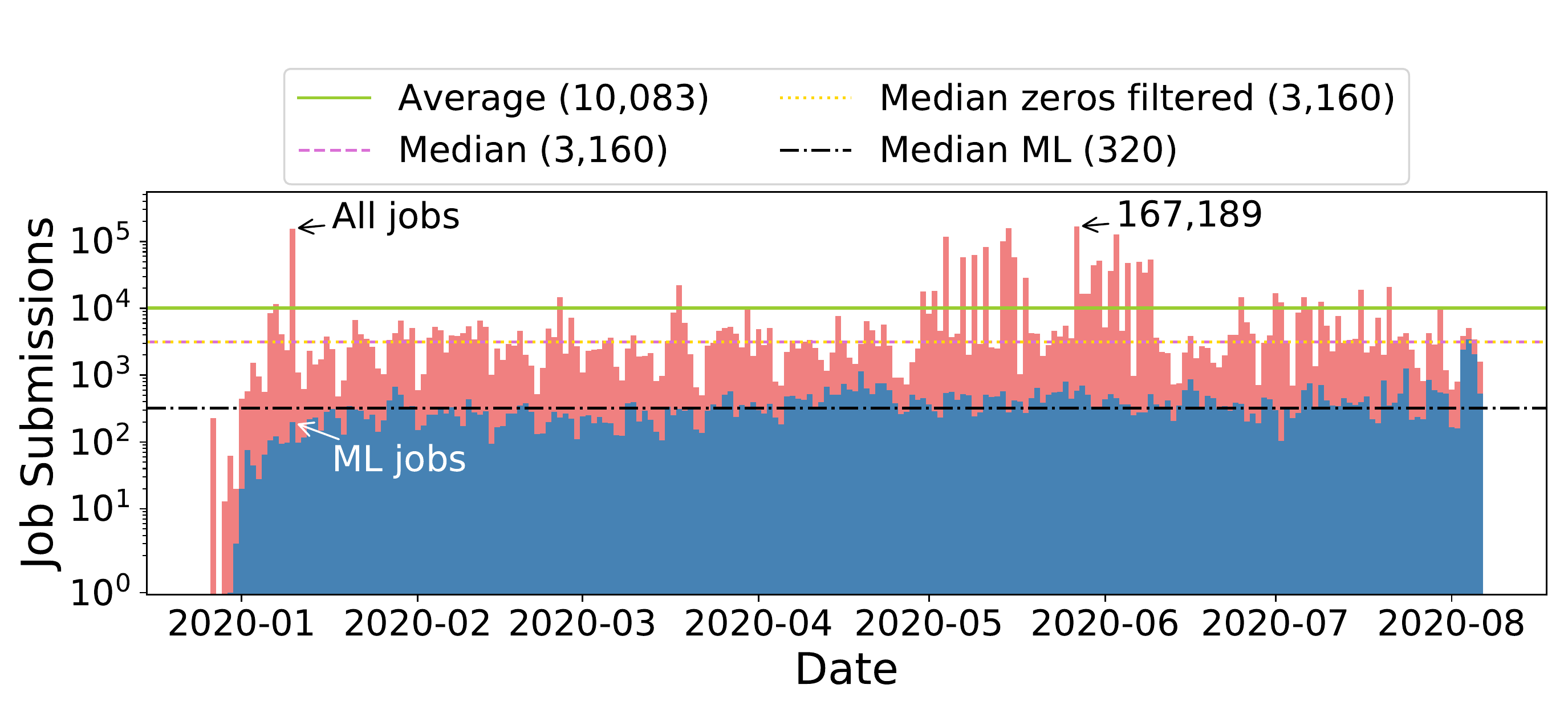}
	\vcutL 
	\vcutM
	\caption{Overview of the daily number of submitted jobs. The maximum is annotated. Logarithmic vertical axis.}
	\label{surfing:fig:characterizations:job-submission-count}
	\vcutM
\end{figure}

\begin{description}
	\mainfinding{surfing:mf:cpu-hours-usage-varies-less}{Arrival and demand are highly variable. The number of submitted jobs per day varies by up to four orders of magnitude. Also, the number of consumed CPU-hours varies by at most two orders of magnitude.}
	\mainfinding{surfing:mf:job-submissions-lisa-diurnal-pattern}{Job submissions have a diurnal (office-like, e.g., 9 to 5) pattern. (See also \refmainfinding{surfing:ssct:characterizations:diurnal1}.)}
	\mainfinding{surfing:mf:lisa-high-job-arrival-rate}{The datacenter has a high job-arrival rate, with several days experiencing over 100,000 job submissions, each.}
	\mainfinding{surfing:mf:lisa-ml-fewer}{Significantly fewer ML jobs are submitted.} %
\end{description}

Combining data depicted in Figures~\ref{surfing:fig:num-jobs-hour-of-day}, \ref{surfing:fig:num-jobs-day-of-week}, and~\ref{surfing:fig:characterizations:job-submission-count}, we observe a highly variable job-arrival process~(\refmainfinding{surfing:mf:cpu-hours-usage-varies-less}).
In contrast, the number of consumed CPU-core hours varies by at most two orders of magnitude, see Figure~\ref{surfing:fig:characterizations:job-squashed-area}.
Unlike the Mustang and OpenTrinity traces, the trace we analyze does feature a clear diurnal pattern in job submissions, depicted in Figure~\ref{surfing:fig:num-jobs-hour-of-day}. 
We observe an office-like daily pattern~(\refmainfinding{surfing:mf:job-submissions-lisa-diurnal-pattern}), with job submissions ramping up in the morning after 9am and lasting until office closing time.
This confirms the expectations of the datacenter operational team.
However, we also observe job-submissions still occur, until 4am.%
Job submissions per day of week also vary greatly, see Figure~\ref{surfing:fig:num-jobs-day-of-week}.
The difference between Sun (lowest, 5,160.2) and Friday (highest, 14,753.9) is $2.86\times$.

Following the method of Amvrosiadis et al.~\cite{DBLP:conf/usenix/AmvrosiadisPGGB18}, we classify as high arrival rate a rate of over 10,000 submitted jobs per day. 
Figure~\ref{surfing:fig:characterizations:job-submission-count} shows the maximum number of submitted jobs on a single day is 167,189, and the average rate is above 10,000~(\refmainfinding{surfing:mf:lisa-high-job-arrival-rate}). 

We observe significantly fewer ML jobs arrive on average, compared to generic jobs~(\refmainfinding{surfing:mf:lisa-ml-fewer}).
Figure~\ref{surfing:fig:characterizations:job-submission-count} depicts this phenomenon. 
The median number of ML-job arrivals per day is only 320, an order of magnitude lower than the median arrival rate for all jobs. 
We link this to the system setup, where users require additional permission to submit jobs to ML nodes.

\subsection{Peak Demand}
\label{surfing:ssct:dataset-outline-resource-utilization}

\begin{figure}[t]
	\centering
		\includegraphics[max width=\linewidth]{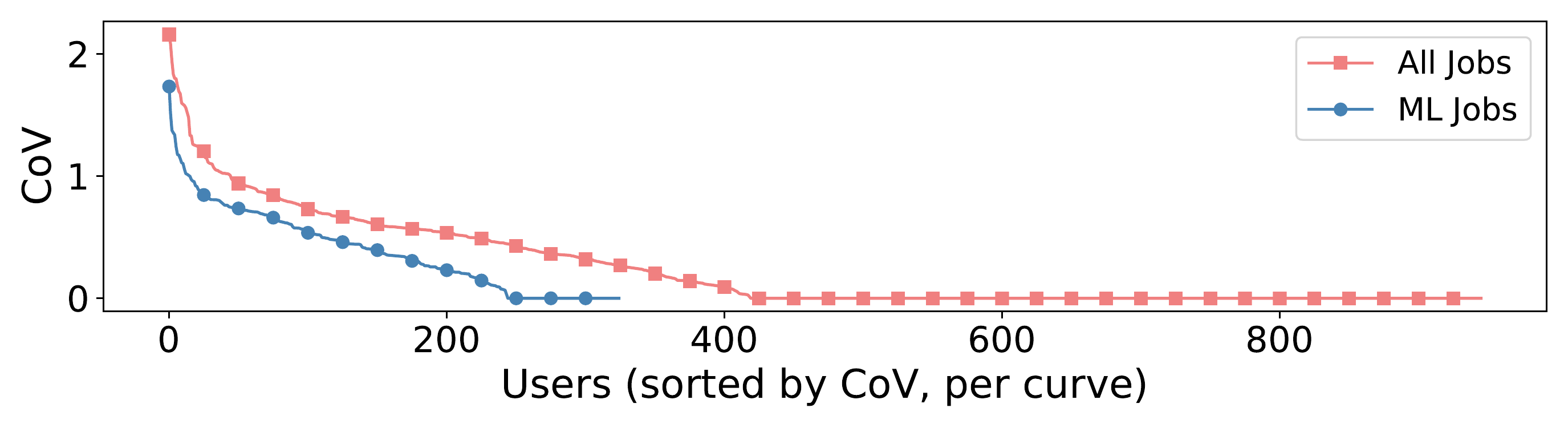}
	\vcutL
	\vcutM
	\caption{CoV of the number of CPU cores requested per user. We show a marker for every 25th user.}
	\label{surfing:fig:cov-cpus-requested-users}
	\vcutM
\end{figure}

\begin{description}
	\mainfinding{surfing:mf:lisa-sub-second-job-arrivals}{There are periods with high,  sub-second job arrivals.}
	\mainfinding{surfing:mf:lisa-user-request-variability-low}{Low variability in the number of requested CPU-cores.}
\end{description}

We analyze now the peak demand of the datacenter. 
The datacenter experiences periods with high, sub-second job-arrival rates~(\refmainfinding{surfing:mf:lisa-sub-second-job-arrivals}); these appear in 
Figure~\ref{surfing:fig:characterizations:job-submission-count} as daily peaks larger than $10^5$.
These translate to resource over-commitment; although the allocation of CPU-cores is limited using cgroups, other resources such as network and disk I/O are not rate-limited. 

Following the approach of Amvrosiadis et al.~\cite{DBLP:conf/usenix/AmvrosiadisPGGB18}, we compute the coefficient of variation~(CoV) of CPU cores requested per user. %
We observe in Figure~\ref{surfing:fig:cov-cpus-requested-users} the CoV is at most 2, with a rapid decrease below 1, low values~(\refmainfinding{surfing:mf:lisa-user-request-variability-low}) similar to those observed at Google as reported by Amvrosiadis et al.

\begin{table}[t] %
	\caption{Fraction of jobs per job state.}
	\vcutM
	\label{surfing:tbl:job-state-fraction}
	\centering
	\adjustbox{max width=\linewidth}{
		\begin{tabular}{@{}rrrrrrrrrr@{}}
			\toprule
			COMPLETED & FAILED & CANCELLED & TIMEOUT & \makecell[l]{OUT OF\\MEMORY} & REQUEUED & \makecell[l]{NODE\\FAILURE} \\ \midrule
			92\% & 6\% & 1\% & 1\% & $<$1\% & $<$1\% & $<$1\% \\ \bottomrule
		\end{tabular}
	}
	\vcutM
\end{table}

\subsection{Failure Analysis}
\label{surfing:ssct:dataset-outline-failure-analysis}

\begin{description}
	\mainfinding{surfing:mf:lisa-low-job-failure-rate}{Most (91.7\%) jobs complete successfully.}
	\mainfinding{surfing:mf:lisa-larger-jobs-lower-success-rate}{Longer-running jobs terminate unsuccessfully more often.}
	\mainfinding{surfing:mf:lisa-unsuccessful-job-consume-significant-resources}{Unsuccessful jobs consume a significant amount of resources, and at worst they do so until they timeout.}
	\mainfinding{surfing:mf:ml-jobs-fail-more-often}{Among all classes of runtimes, ML jobs terminate unsuccessfully more often than other jobs.}

\end{description}

Relatively few jobs have unsuccessful job outcomes, see Table~\ref{surfing:tbl:job-state-fraction}.
As we observe, more than 91\% of jobs complete successfully~(\refmainfinding{surfing:mf:lisa-low-job-failure-rate}), which is more than the highest fraction reported by Amvrosiadis et al.~\cite{DBLP:conf/usenix/AmvrosiadisPGGB18}.

In contrast, we observe that longer jobs and jobs that consume more resources tend to fail more often~(\refmainfinding{surfing:mf:lisa-larger-jobs-lower-success-rate}), see Figure~\ref{surfing:fig:job-outcome-per-runtime-category}.
For the latter category, for all (ML) jobs, a high fraction of 51.2\% (55.8\%) of the runtime is spent on non-completed jobs.
For long-running jobs, (ML) jobs that do not complete consume 13.8\% (51.9\%)~(\refmainfinding{surfing:mf:lisa-unsuccessful-job-consume-significant-resources}).

Across all job durations, between 32.3\% and 55.8\% of the ML jobs complete unsuccessfully; this is more often than all jobs (12.9-51.2\%)~(\refmainfinding{surfing:mf:ml-jobs-fail-more-often}).
We depict the total sum of job runtimes and their fraction per job state. The behavior of longer jobs failing more often is mainly due to timeouts, as there is a 5-day limit in the datacenter, as the operators reported. The data shows clearly that larger jobs fail more often and consume more time than smaller jobs.

We have presented an in-depth analysis of several of the metrics listed in the archive we consider. We continue by presenting how these results can be leveraged by the community at large to better understand datacenter behavior, how to build more efficient datacenters and how to predict the artifacts of their operations.

%% file: 07_implications.tex
\section{Implications of Our Results}
\label{surfing:sct:implications}\label{surfing:sec:implications}

For the principle of
holistic analysis to gain traction, 
the community needs to find useful guidelines and applications. 
Toward this end, but limited in scope, 
this section presents several examples.

\subsection{High-Frequency Data for Prediction (\resqref{wl:lstm})}\label{surfing:sec:implications:prediction}

\begin{figure}[t]
	\centering
	\includegraphics[max width=\linewidth]{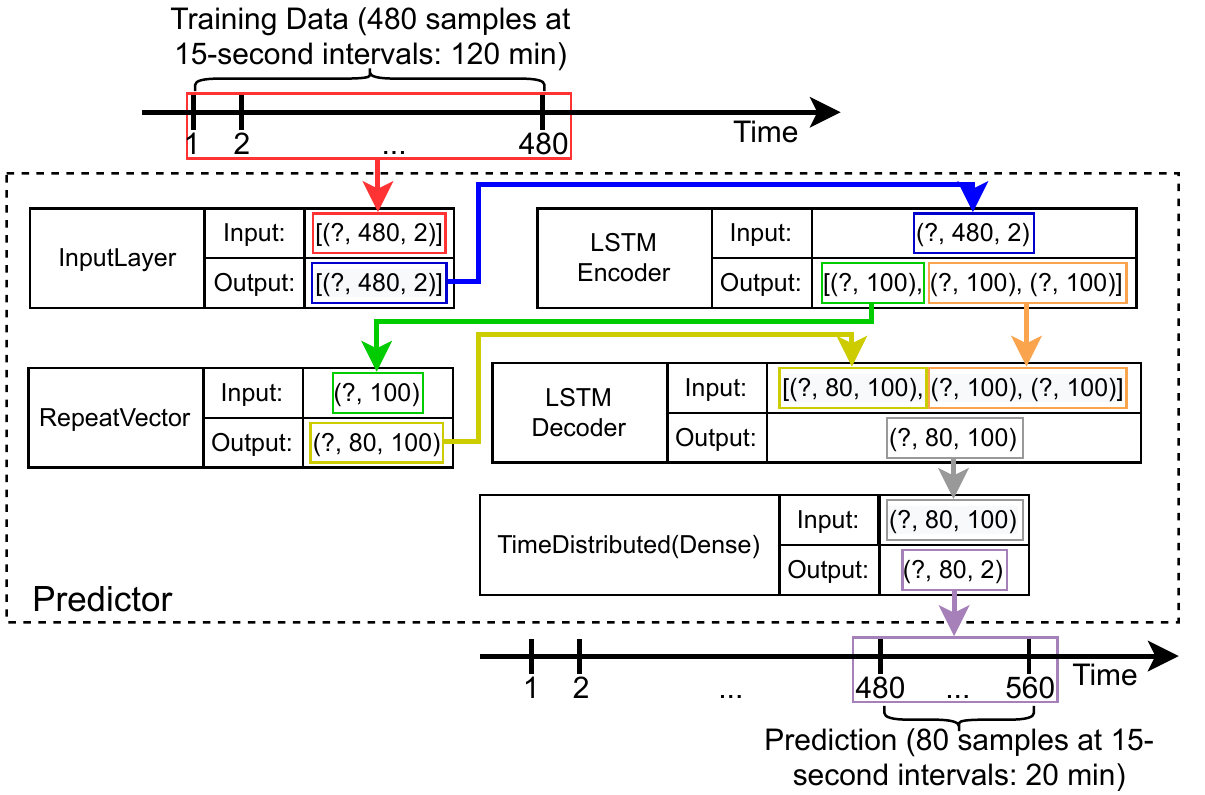}
	\vcutL
	\vcutM
	\caption{Architecture of the LSTM model. (480 input items)}
	\label{surfing:fig:anomaly-detection:lstm}
	\vcutL
\end{figure}

Online performance prediction is a well-established instrument for datacenter operation, useful among others for optimization at both system~\cite{DBLP:conf/sc/SwanyW02} and application-level~\cite{DBLP:conf/ispass/LiMSSSM09}. In the past two decades, online performance prediction using machine learning~(ML) has become common~\cite{DBLP:journals/pe/DobberMK07}. Because ML methods such as Long Short-Term Memory (LSTM) networks~\cite{DBLP:conf/hpdc/Baughman0WFC18} can operate on any data, new questions arise: 
\textit{Is higher-frequency data more useful for online prediction than low-frequency data? How high should the frequency be?}

To address this question, we focus in this section on the use of LSTM to predict online performance data for the next 20 minutes of operation. LSTM is common in practice, including for datacenter workloads~\cite{DBLP:conf/bigdataconf/Siami-NaminiTN19}, which allows us to focus on the new research questions. 
We vary the granularity of the input, from high frequency~(15\,seconds) to low (up to 10\,minutes), and observe the quality of the prediction provided by the LSTM predictor (using a common metric, Huber loss). Common practice in datacenters and public clouds samples at 5-/10-minutes, and even lower frequency.

We employ the LSTM model depicted in Figure~\ref{surfing:fig:anomaly-detection:lstm}. Using it requires two setup elements. First, to select prediction metrics and prepare data for LSTM use, we focus on data studied during pair-wise correlation analysis (in Section~\ref{surfing:sct:correlations}). We select \texttt{node\_load1} 
and \texttt{node\_sockstat\_sockets\_used}, which we find to correlate well. 
We normalize data to make it suitable as LSTM input.

Second, to train the LSTM model for prediction (inference), we create %
four different datasets from the same data: (1) the \enquote{raw} 15-second interval data as present in the original SURFace dataset, and data resampled in (2) 1-minute, (3) 5-minute, and (4) 10-minute time intervals.
We train the model on each dataset, resulting in four separate predictors.
We split each training dataset into 2-hour chunks; as Figure~\ref{surfing:fig:anomaly-detection:lstm} depicts, for the 15-second dataset this results in 480 input tuples. %
(The three other datasets include 120, 24, and 12 inputs, respectively.)
We configure the model to generate a prediction for 20-minute window, with predictions 15\,seconds apart (80\,predictions). 
We use 10\% of the entire data for evaluation purposes; this allows to evaluate the generality of the model, by inspecting its performance on unseen data.

\begin{table}[t]
	\caption{Loss values (Huber) for different nodes per time granularity. Lower values are better. Aggregation at 5-min or higher are what datacenter operators commonly use.}
	\small
	\label{surfing:tbl:loss-value-LSTM}
	\vcutM
	\centering
	\begin{tabular}{@{}lrrrr@{}}
\toprule
\textbf{Node} & \textbf{15-sec}                          & \textbf{1-min}                           & \textbf{5-min}                           & \textbf{10-min} \\ \midrule
r10n13        & \cellcolor[HTML]{C0C0C0}\textbf{0.0096} & 0.0100                                  & 0.0100                                  & 0.0163         \\
r11n18        & 0.0175                                  & \cellcolor[HTML]{C0C0C0}\textbf{0.0169} & 0.0170                                  & 0.0261         \\
r12n6         & 0.0175                                  & \cellcolor[HTML]{C0C0C0}\textbf{0.0165} & 0.0151                                  & 0.0266         \\
r13n23        & \cellcolor[HTML]{C0C0C0}\textbf{0.0102} & \cellcolor[HTML]{C0C0C0}\textbf{0.0102} & 0.0113                                  & 0.0190         \\
r14n2         & 0.0461                                  & \cellcolor[HTML]{C0C0C0}\textbf{0.0311} & 0.0401                                  & 0.0560         \\
r15n23        & 0.0185                                  & 0.0182                                  & \cellcolor[HTML]{C0C0C0}\textbf{0.0175} & 0.0285         \\
r25n32        & 0.0206                                  & 0.0215                                  & \cellcolor[HTML]{C0C0C0}\textbf{0.0201} & 0.0308         \\
r26n8         & \cellcolor[HTML]{C0C0C0}\textbf{0.0117} & 0.0118                                  & \cellcolor[HTML]{C0C0C0}\textbf{0.0117} & 0.0172         \\
r27n16        & \cellcolor[HTML]{C0C0C0}\textbf{0.0222} & 0.0234                                  & 0.0276                                  & 0.0320         \\
r29n3         & \cellcolor[HTML]{C0C0C0}\textbf{0.0026} & 0.0034                                  & 0.0030                                  & 0.0044         \\
r30n6         & 0.0330                                  & \cellcolor[HTML]{C0C0C0}\textbf{0.0319} & \cellcolor[HTML]{C0C0C0}\textbf{0.0319} & 0.0477         \\
r31n5         & \cellcolor[HTML]{C0C0C0}\textbf{0.0249} & 0.0300                                  & 0.0268                                  & 0.0362         \\
r38n1         & \cellcolor[HTML]{C0C0C0}\textbf{0.0027} & 0.0029                                  & 0.0036                                  & 0.0097         \\ 
\bottomrule
\end{tabular}
	\vcutL
\end{table}

Table~\ref{surfing:tbl:loss-value-LSTM} compares the loss values for the four trained models, with highlighted cells indicating which model delivers the best prediction. 
We perform this analysis on a randomly selected set of 13 general and ML nodes.
We show here results for 5 nodes; these are representative for all the results in this analysis. %
The LSTM model trained with the 10-minute dataset is never the best predictor, indicating datacenter operators should use higher-frequency data for prediction. %
Among the 15-second and 1-/5-minute datasets, we find the former is the best-performer 7 times, with 3 ties, suggesting choice in 
the accuracy-performance trade-off. 
To conclude: \textit{Higher-frequency metric data improves performance predictions when using LSTMs. We recommend datacenters collect and provide such data.}

\subsection{Actionable Insights (\resqref{wl:storagecomp}, \resqref{wl:corr})}\label{surfing:sec:implications:actionable}\label{surfing:sec:implications:overhead}

\textbf{Computation and Storage Overheads.} The computational load for training LSTM models on 1- or 5-minute data is significantly lower than for 15-second intervals; applied to 2-hour chunks, we find all training remains feasible at datacenter-level, and inference imposes a negligible computational cost.
Additionally, the amount of storage required for the fine-grained data is non-linear with the number of samples due to compression. Intuitively, storing a $2\times$ larger dataset would require $2\times$ storage. However, with modern storage formats, that leverage compression and columnar formats, this is not the case. Using \textit{snappy} compression (the default for \textit{parquet}), the data representing a 10-minute granularity snapshot for the two metrics used in this example requires 32.77\,MB of storage. In turn, only 277.56\,MB is required to store data with a granularity of 15 seconds, 
so increasing the volume of uncompressed data $40\times$ increases the actual storage by only $8.47\times$.
Therefore, leveraging modern data storage techniques enables storing high-frequency data with sub-linear overheads. To conclude, \textit{higher-frequency metric data incurs both computational and storage overheads, but these seem worthwhile when compared with the benefits they enable.} %

\textbf{Metric Correlations.} The analysis we depict in Figure~\ref{surfing:fig:correlation-metrics-50-days-0.9} shows a novel insight: only 40 pairs of low-level metrics are correlated over a time period of 50 days. The pairs of correlated metrics differ significantly per day, leading to over 14,000 \textit{unique} correlations over the entire period. 
Having so many pairs that correlate infrequently shows strong evidence that correlations are workload dependent, therefore \textit{as many metrics must be captured as frequently as possible}. 
Our guideline is to \textit{only eliminate the metrics that are correlated over long periods of time.} 

\begin{table}[]
\centering
\small
\caption{Correlated metrics identified by analyzing the dataset generated by analysis in Figure~\ref{surfing:fig:correlation-metrics-50-days-0.9}.}
\label{tab:anecdotal_correlations}
\vcutM
\begin{tabular}{@{}ll@{}}
\toprule
Metric 1                   & Metric 2             \\ \midrule
server swap memory         & GPU temperature      \\
network receive fifo    & GPU temperature \\
TCP open sockets           & GPU temperature      \\ \bottomrule
\end{tabular}
\vcutL
\end{table}

We conclude this section with anecdotal insight from our correlation analysis. We find many metrics that, intuitively, correlate persistently: \textit{load1} with \textit{load15}, \textit{netstat TCP data} with 
\textit{netstat IP data}, and \textit{swap memory} with \textit{free memory}. 
By manually inspecting the correlations that are not persistent over time, we find other, more interesting correlations that would be difficult to predict even by experts. 
Table~\ref{tab:anecdotal_correlations} presents 
three %
metrics linking IO and GPU processing, corroborating recent ML benchmarks~\cite{jansen2020ddlbench}. 
\textit{Systematic insight into (multi-)metric correlations remains an open challenge.}

\subsection{Fine-grained Data in Design, Tuning (\resqref{wl:implications})}
\label{surfing:sec:implications:design}

Our final guideline is to \textit{use fine-grained data for designing and tuning datacenters, from individual chips to full-system procurement.} 
We support this guideline with qualitative analysis.

Datacenters are often acquiring homogeneous batches of hardware. Often, for datacenters for scientific computing and engineering, nodes pack a large x86 CPU and large amounts of memory. Clusters equipped for HPC and machine learning often also add GPUs and high-speed interconnects. The power envelope of datacenters has constantly increased, and modern large-scale datacenters approach the limits of what our society can leverage in terms of power while being mindful of carbon emissions~\cite{science/Masanet+20,DutchDataCenters20}. Others have considered power savings by means of reducing cooling~\cite{el2012temperature}, but that is only one example of the many aspects that could be considered. In this paper, we have analyzed many metrics, all with potential impact on how datacenters could be tuned and designed. 
We posit that using such data for customizing datacenters suited to their user's needs is key for efficiency. 
Using the resource usage profiles uncovered in this work one could, for example, build machine-learning clusters more efficiently by leveraging lower-power CPUs (e.g., ARM and RISC-V) next to power-hungry GPUs. In GPU-based ML workloads, power-hungry x86 CPUs are underutilized, being mostly used in data pre-processing and data movement.
Moreover, as memory usage is low in our traces, for similar workloads the designer need not to purchase large amounts of RAM.
For inadvertent peak-loads, designers could leverage software disaggregation methods~\cite{gu2017efficient,uta2016towards}, instead of %
hardware acquisition.

Uncovering inefficiencies in datacenters by holistic performance analysis approaches can also lead to improved chip design. In the post-Moore era, this is an avenue beginning to be explored by large tech companies and hardware manufacturers. Google pioneered %
optimizing ML training with TPUs~\cite{jouppi2018motivation}. This trend continues at organizations %
like Amazon or Nvidia, who  %
are building inference-tailored chips~\cite{Chips/Amazon,Chips/NVIDIA}, %
or even deep-learning programmable engines~\cite{Chips/XILINX}. %
Only with such analysis, practitioners could tackle both performance, power consumption and other important metrics at the same time. Similar trends have already started at the network level, where in-network computing is already a reality~\cite{stephens2018your}. Significantly improving network performance, and releasing pressure from CPUs is something that our data already supports (see Figures~\ref{surfing:fig:characterizations:general-resource-usage} and~\ref{surfing:fig:characterizations:network-analysis:transmitted-packets-job-time-scatterplot}).
\addition{Already, the analysis we have conducted in this work has helped the datacenter operator improve the design of their next monitoring system.}

%% file: 09_related_work.tex
\section{Related Work}
\label{surfing:sct:related-work}

In this section, we group related works by topic and discuss our contributions with respect to each.
Overall, we propose a holistic method of analysis, and 
use it on a dataset with unprecedented temporal and spatial granularity among public datasets.

\textbf{Datacenter operations:}
Several articles provide a holistic view of datacenter operations, including job allocation~\cite{DBLP:conf/sc/AndreadisVMI18}, cloud services~\cite{liu2011nist}, physical network~\cite{lam2010fiber}, etc.
Different from related work, our article provides a view of the effect of the workload on machine metrics.
This complements prior work and aids in understanding the operations within modern datacenters.

\textbf{Characterizations of workloads:}
There are various articles on the topic of characterizing workloads from Google~\cite{DBLP:conf/ccgrid/RosaCB15, DBLP:journals/tsc/RosaCB17}, FinTech~\cite{ DBLP:conf/usenix/AmvrosiadisPGGB18, DBLP:journals/tpds/VersluisMTHPDI20}, scientific computing environments~\cite{DBLP:conf/usenix/AmvrosiadisPGGB18, DBLP:journals/tpds/VersluisMTHPDI20,ferreiradasilva2020works}, etc.
Adding to this topic, we demonstrate our workload is unique in terms of properties.
Additionally, many of the jobs are from the ML domain, which, combined with the machine metric characterization, provides interesting (and sometimes contrasting) insights.

\textbf{Characterizations of machine metrics:}
There is also a body of related work focusing on machine metrics.
Closest to this work is the work done on a subset of the dataset we characterize~\cite{DBLP:journals/usenix-login/UtaLIMPC20}.
Other related work focuses on few, high-level metrics~\cite{DBLP:conf/date/KimRAL13, DBLP:conf/ccgrid/ShenBI15, DBLP:conf/IEEEcloud/BirkeCS12}.
Different in our work is the various additional and novel angles we use and the discussion of the implications in various directions such as hardware design and societal aspects.

\textbf{Metric correlations:}
Some related works make use of correlation coefficients.
There are many applications to finding correlations, e.g., finding metric correlations that hold over longer periods of time~\cite{DBLP:conf/sosp/CortezBMRFB17}, finding (virtual) machines executing the same application~\cite{canali2017identifying}, finding (virtual) machines that correlate in resource utilization~\cite{DBLP:conf/date/KimRAL13} to minimize contention, checking for correlations between resources requested in datacenters~\cite{DBLP:conf/ccgrid/ShenBI15}, etc.
We focus not only on finding correlations that hold over a long period of time, but also demonstrate that correlations are workload dependent. We use three correlation methods to study how metric pairs are correlated, and use significantly more metric pairs (over 14,000).

%% file: 10_conclusion.tex
\section{Conclusion and Future Directions}
\label{surfing:sct:conclusion}

For decades we have been focusing on optimizing systems for only the metrics we measure.
Thus, to conquer the ever-increasing complexity of our datacenters, we posit the need for holistic overview of such systems. 
In this work we propose a holistic method of analysis for datacenter operations.

We applied the method on a public, long-term datacenter traces of unprecedented temporal and spatial granularity. 
Poring billions of data points in total, with samples collected at 15-second intervals covering hundreds of operational metrics, we characterized the machines, power consumption, and workload. We distinguished between generic and ML-specific information, and between regular operation and operation during the 2020 COVID-19 pandemic. 
We made over 30 main observations, which give detailed, holistic insight into the operation of a public scientific infrastructure.
Finally, we discussed the implications of our findings on online ML-based prediction, and on long-term datacenter design and tuning.

We envision our work, and similar pioneering efforts, as motivators for a community-driven approach embracing holistic analysis. %
Concretely, we envision our analysis will show organizations the potential of collecting, and ultimately releasing, many more fine-grained datasets.
We also envision studies comparing such datasets, finding invariants and trends, and thus bolstering fundamental knowledge in the field.
Last, we envision new techniques for datacenter operations, from dynamic scheduling to long-term resource procurement, all enhanced by the use of holistic data and considerations.

For future work, investigating different forecasting techniques can be interesting to see what leverage this fine grained data can further offer.
Due to the richness of this dataset, we believe more interesting characterizations can be done, which is another item for future work.
Comparing the findings of this -- and potentially followup -- work with another rich dataset would be very interesting to observe if findings hold across multiple systems and workloads.
Such comparisons can lead to the development of new scheduling approaches or the design of new systems.
Furthermore, having access to multi-year data can eliminate the effect of seasonality as discussed in our threats to validity.
Accounting for, eliminating, and comparing the effect of seasonality will further contribute to our understanding of these systems.

%% file: main.bbl
\begin{thebibliography}{10}

\bibitem{Chips/Amazon}
{Amazon, Inc.}
\newblock {AWS Inferentia: High performance machine learning inference chip,
  custom designed by AWS}.
\newblock \url{https://aws.amazon.com/machine-learning/inferentia/}, 2018-2021.

\bibitem{DBLP:conf/usenix/AmvrosiadisPGGB18}
Amvrosiadis et~al.
\newblock On the diversity of cluster workloads and its impact on research
  results.
\newblock In {\em ATC}, 2018.

\bibitem{DBLP:conf/sc/AndreadisVMI18}
Andreadis et~al.
\newblock A reference architecture for datacenter scheduling: design,
  validation, and experiments.
\newblock In {\em SC}, 2018.

\bibitem{DBLP:conf/hpdc/Baughman0WFC18}
Baughman et~al.
\newblock Predicting amazon spot prices with {LSTM} networks.
\newblock In {\em {ScienceCloud@HPDC}}, pages 1:1--1:7, 2018.

\bibitem{DBLP:conf/IEEEcloud/BirkeCS12}
Birke et~al.
\newblock Data centers in the cloud: {A} large scale performance study.
\newblock In {\em CLOUD}, 2012.

\bibitem{DBLP:conf/icppw/BourassaJBCJVS19}
Bourassa et~al.
\newblock Operational data analytics: Optimizing the national energy research
  scientific computing center cooling systems.
\newblock In {\em {ICPP Workshop}}, 2019.

\bibitem{DBLP:journals/cacm/BouwersVD12}
Bouwers et~al.
\newblock Getting what you measure.
\newblock {\em CACM}, 55, 2012.

\bibitem{canali2017identifying}
Canali and Lancellotti.
\newblock Identifying communication patterns between virtual machines in
  software-defined data centers.
\newblock {\em SIGMETRICS}, 44, 2017.

\bibitem{DBLP:conf/ipps/ChapinCFJLSST99}
Chapin, Cirne, Feitelson, et~al.
\newblock Benchmarks and standards for the evaluation of parallel job
  schedulers.
\newblock In {\em {JSSPP}}, pages 67--90, 1999.

\bibitem{DBLP:conf/sosp/CortezBMRFB17}
Cortez et~al.
\newblock Resource central: Understanding and predicting workloads for improved
  resource management in large cloud platforms.
\newblock In {\em SOSP}, 2017.

\bibitem{dayarathna2015data}
Dayarathna et~al.
\newblock Data center energy consumption modeling: A survey.
\newblock {\em COMST}, 18, 2015.

\bibitem{dean2013tail}
Dean and Barroso.
\newblock The tail at scale.
\newblock {\em CACM}, 56, 2013.

\bibitem{DBLP:journals/pe/DobberMK07}
Dobber et~al.
\newblock A prediction method for job runtimes on shared processors: Survey,
  statistical analysis and new avenues.
\newblock {\em Perf. Eval.}, 64, 2007.

\bibitem{DutchDataCenters20}
{Dutch Data Center Association}.
\newblock {State of the Dutch data centers}.
\newblock
  \url{https://www.dutchdatacenters.nl/en/publications/state-of-the-dutch-data-centers-2020/},
  2020.

\bibitem{corona}
{Dutch Government}.
\newblock {Ontwikkeling COVID-19 in grafieken}.
\newblock \url{https://www.rivm.nl/coronavirus-covid-19/grafieken}, 2020.

\bibitem{duy2010performance}
Duy et~al.
\newblock Performance evaluation of a green scheduling algorithm for energy
  savings in cloud computing.
\newblock In {\em IPDPSW}, 2010.

\bibitem{el2012temperature}
El-Sayed et~al.
\newblock Temperature management in data centers: why some (might) like it hot.
\newblock In {\em SIGMETRICS}, 2012.

\bibitem{DBLP:journals/jpdc/FeitelsonTK14}
Feitelson et~al.
\newblock Experience with using the parallel workloads archive.
\newblock {\em JDPC}, 74, 2014.

\bibitem{DBLP:conf/ispdc/GhiasvandC19}
Ghiasvand and Ciorba.
\newblock Anomaly detection in high performance computers: {A} vicinity
  perspective.
\newblock In {\em {ISPDC}}, pages 112--120, 2019.

\bibitem{ghodsi2011dominant}
Ghodsi et~al.
\newblock Dominant resource fairness: Fair allocation of multiple resource
  types.
\newblock In {\em NSDI}, volume~11, 2011.

\bibitem{goiri2011greenslot}
Goiri et~al.
\newblock Greenslot: scheduling energy consumption in green datacenters.
\newblock In {\em SC}, 2011.

\bibitem{BrendanGreggLoad1}
Gregg.
\newblock {Linux load averages: Solving the Mystery}.
\newblock
  \url{http://www.brendangregg.com/blog/2017-08-08/linux-load-averages.html},
  2017.

\bibitem{gu2017efficient}
Gu et~al.
\newblock Efficient memory disaggregation with infiniswap.
\newblock In {\em NSDI}, 2017.

\bibitem{gunawi2018fail}
Gunawi et~al.
\newblock Fail-slow at scale: Evidence of hardware performance faults in large
  production systems.
\newblock {\em TOS}, 14, 2018.

\bibitem{DBLP:journals/csur/IbidunmoyeHE15}
Ibidunmoye, Hern{\'{a}}ndez{-}Rodriguez, and Elmroth.
\newblock Performance anomaly detection and bottleneck identification.
\newblock {\em {ACM} Comput. Surv.}, 48(1):4:1--35, 2015.

\bibitem{DBLP:journals/fgcs/IosupLJADWE08}
Iosup et~al.
\newblock The grid workloads archive.
\newblock {\em FGCS}, 24, 2008.

\bibitem{jansen2020ddlbench}
Jansen et~al.
\newblock Ddlbench: Towards a scalable benchmarking infrastructure for
  distributed deep learning.
\newblock In {\em DLS at ICS}, pages 31--39, 2020.

\bibitem{mt-ml3}
Jeon et~al.
\newblock Analysis of large-scale multi-tenant gpu clusters for dnn training
  workloads.
\newblock In {\em ATC}, 2019.

\bibitem{jouppi2018motivation}
Jouppi et~al.
\newblock Motivation for and evaluation of the first tensor processing unit.
\newblock {\em IEEE Micro}, 38(3):10--19, 2018.

\bibitem{DBLP:conf/date/KimRAL13}
Kim et~al.
\newblock Correlation-aware virtual machine allocation for energy-efficient
  datacenters.
\newblock In {\em DATE}, 2013.

\bibitem{science/Masanet+20}
Koomey et~al.
\newblock Recalibrating global data center energy-use estimates.
\newblock {\em Science}, 367, 2020.

\bibitem{lam2010fiber}
Lam et~al.
\newblock Fiber optic communication technologies: What's needed for datacenter
  network operations.
\newblock {\em IEEE Communications Magazine}, 48, 2010.

\bibitem{laursen_olason_kristian_valur_2020_3878143}
Laursen et~al.
\newblock {Beneath the SURFace: An MRI-like View into the Life of a 21st
  Century Datacenter}, 2020.
\newblock Zenodo dataset, \url{https://zenodo.org/record/3878143}.

\bibitem{DBLP:journals/cacm/LegrandVCGBC09}
Legrand et~al.
\newblock Monitoring and control of large systems with monalisa.
\newblock {\em Commun. {ACM}}, 52(9):49--55, 2009.

\bibitem{mt-ml1}
Li et~al.
\newblock Ease.ml: Towards multi-tenant resource sharing for machine learning
  workloads.
\newblock {\em Proc. VLDB Endow.}, 11, 2018.

\bibitem{DBLP:conf/ispass/LiMSSSM09}
Jiangtian Li et~al.
\newblock Machine learning based online performance prediction for runtime
  parallelization and task scheduling.
\newblock In {\em {ISPASS}}, pages 89--100, 2009.

\bibitem{liu2011nist}
Liu et~al.
\newblock Nist cloud computing reference architecture.
\newblock {\em NIST special publication}, 500, 2011.

\bibitem{lockwood2018year}
Lockwood et~al.
\newblock A year in the life of a parallel file system.
\newblock In {\em SC}, 2018.

\bibitem{maricq2018taming}
Maricq et~al.
\newblock Taming performance variability.
\newblock In {\em OSDI}, 2018.

\bibitem{DBLP:conf/hpdc/NettiMGOTO020}
Netti et~al.
\newblock {DCDB} wintermute: Enabling online and holistic operational data
  analytics on {HPC} systems.
\newblock In {\em {HPDC}}, 2020.

\bibitem{Chips/NVIDIA}
{NVIDIA}.
\newblock {NVIDIA Deep Learning Accelerator (NVDLA)}.
\newblock \url{http://nvdla.org/}, 2017-2021.

\bibitem{DBLP:journals/cacm/Ousterhout18}
Ousterhout.
\newblock Always measure one level deeper.
\newblock {\em CACM}, 61, 2018.

\bibitem{DBLP:conf/sc/PatelLKRAT20}
Patel et~al.
\newblock Job characteristics on large-scale systems: long-term analysis,
  quantification, and implications.
\newblock In {\em {SC}}, 2020.

\bibitem{DBLP:conf/ipps/PatelWEHZT20}
Patel et~al.
\newblock What does power consumption behavior of {HPC} jobs reveal? :
  Demystifying, quantifying, and predicting power consumption characteristics.
\newblock In {\em IPDPS}, 2020.

\bibitem{pedram2012energy}
Pedram.
\newblock Energy-efficient datacenters.
\newblock {\em TCAD}, 31, 2012.

\bibitem{DBLP:conf/ccgrid/RosaCB15}
Ros{\`{a}} et~al.
\newblock Predicting and mitigating jobs failures in big data clusters.
\newblock In {\em CCGrid}, 2015.

\bibitem{DBLP:journals/tsc/RosaCB17}
Ros{\`{a}} et~al.
\newblock Failure analysis and prediction for big-data systems.
\newblock {\em TSC}, 10, 2017.

\bibitem{schober2018correlation}
Schober et~al.
\newblock Correlation coefficients: appropriate use and interpretation.
\newblock {\em Anesthesia \& Analgesia}, 126, 2018.

\bibitem{DBLP:conf/usenix/ShahradFGCBCLTR20}
Shahrad et~al.
\newblock Serverless in the wild: Characterizing and optimizing the serverless
  workload at a large cloud provider.
\newblock In {\em ATC}, 2020.

\bibitem{DBLP:conf/ccgrid/ShenBI15}
Shen et~al.
\newblock Statistical characterization of business-critical workloads hosted in
  cloud datacenters.
\newblock In {\em CCGrid}, 2015.

\bibitem{DBLP:conf/bigdataconf/Siami-NaminiTN19}
Siami{-}Namini et~al.
\newblock The performance of {LSTM} and bilstm in forecasting time series.
\newblock In {\em {IEEE} Big Data}, pages 3285--3292, 2019.

\bibitem{2010-dapper}
Sigelman et~al.
\newblock Dapper, a large-scale distributed systems tracing infrastructure.
\newblock 2010.

\bibitem{DBLP:conf/eScience/SilvaCJVD14}
Silva et~al.
\newblock Community resources for enabling research in distributed scientific
  workflows.
\newblock In {\em e-Science}, 2014.

\bibitem{ferreiradasilva2020works}
Silva et~al.
\newblock Workflowhub: Community framework for enabling scientific workflow
  research and development.
\newblock In {\em WORKS}, 2020.

\bibitem{stephens2018your}
Stephens et~al.
\newblock Your programmable nic should be a programmable switch.
\newblock In {\em HotNets}, 2018.

\bibitem{DBLP:conf/sc/SwanyW02}
Swany and Wolski.
\newblock Multivariate resource performance forecasting in the network weather
  service.
\newblock In {\em {SC}}, pages 30:1--10, 2002.

\bibitem{thain2005distributed}
Thain et~al.
\newblock Distributed computing in practice: the condor experience.
\newblock {\em CCPE}, 17, 2005.

\bibitem{tirmazi2020borg}
Tirmazi et~al.
\newblock Borg: the next generation.
\newblock In {\em EuroSys}, 2020.

\bibitem{uta2016towards}
Uta et~al.
\newblock Towards resource disaggregation—memory scavenging for scientific
  workloads.
\newblock In {\em CLUSTER}, 2016.

\bibitem{DBLP:journals/usenix-login/UtaLIMPC20}
Uta et~al.
\newblock Beneath the surface: An mri-like view into the life of a 21st-century
  datacenter.
\newblock {\em USENIX ;login:}, 45, 2020.

\bibitem{uta2020big}
Uta et~al.
\newblock Is big data performance reproducible in modern cloud networks?
\newblock In {\em NSDI}, 2020.

\bibitem{DBLP:conf/sc/VazhkudaiMTZWOG17}
Vazhkudai et~al.
\newblock {GUIDE:} a scalable information directory service to collect,
  federate, and analyze logs for operational insights into a leadership {HPC}
  facility.
\newblock In {\em {SC}}, 2017.

\bibitem{verma2012two}
Verma et~al.
\newblock Two sides of a coin: Optimizing the schedule of mapreduce jobs to
  minimize their makespan and improve cluster performance.
\newblock In {\em MASCOTS}, 2012.

\bibitem{verma2015large}
Verma et~al.
\newblock Large-scale cluster management at google with borg.
\newblock In {\em EuroSys}, 2015.

\bibitem{DBLP:journals/tpds/VersluisMTHPDI20}
Versluis et~al.
\newblock The workflow trace archive: Open-access data from public and private
  computing infrastructures.
\newblock {\em TPDS}, 31, 2020.

\bibitem{data:FAIR16}
Wilkinson et~al.
\newblock {The FAIR Guiding Principles for scientific data management and
  stewardship}.
\newblock {\em Nature {SciData}}, 3, 2016.

\bibitem{DBLP:journals/concurrency/XiaoYER16}
Xiao et~al.
\newblock Using spearman's correlation coefficients for exploratory data
  analysis on big dataset.
\newblock {\em CCPE}, 28, 2016.

\bibitem{Chips/XILINX}
{XILINX}.
\newblock {The Xilinx® Deep Learning Processor Unit (DPU)}.
\newblock \url{https://www.xilinx.com/products/intellectual-property/dpu.html},
  2020-2021.

\bibitem{DBLP:conf/wosp/XiongPZG13}
Xiong et~al.
\newblock vperfguard: an automated model-driven framework for application
  performance diagnosis in consolidated cloud environments.
\newblock In {\em {ICPE}}, pages 271--282, 2013.

\bibitem{zhang2018comparison}
Zhang et~al.
\newblock Comparison and evaluation of air cooling and water cooling in
  resource consumption and economic performance.
\newblock {\em Energy}, 2018.

\bibitem{2016-osdi-profiling}
Zhao et~al.
\newblock Non-intrusive performance profiling for entire software stacks based
  on the flow reconstruction principle.
\newblock In {\em OSDI}, 2016.

\end{thebibliography}
